\documentclass[titlepage,12pt]{utarticle}

\usepackage{amsmath,amssymb,amsthm,amscd,latexsym}

\usepackage{color}
\definecolor{dblue}{rgb}{0,0,.7}
\definecolor{indigo}{RGB}{50,0,105}

\usepackage[hyperindex=true, colorlinks=true, pagecolor=dblue, bookmarksopen=false, urlcolor=dblue, linktocpage,linkcolor=dblue, citecolor=indigo]{hyperref}      

\usepackage[all]{xy}
\CompileMatrices

\newtheorem{thm}{Theorem}[section]
\newtheorem{lemma}[thm]{Lemma}

\newtheorem{prop}[thm]{Proposition}

\def\roof{{\mbox{\tiny \mbox{$\vee$}}}}
\def\comp{\mbox{\scriptsize \mbox{$\,\circ\, $}}}
\def\poso#1{#1\save="x"!LD+<0pt,-0.5mm>;  "x"!RD+<0pt,-0.5mm>**\dir{.}\restore}

\newdimen\tableauside\tableauside=1.0ex
\newdimen\tableaurule\tableaurule=0.4pt
\newdimen\tableaustep
\def\phantomhrule#1{\hbox{\vbox to0pt{\hrule height\tableaurule width#1\vss}}}
\def\phantomvrule#1{\vbox{\hbox to0pt{\vrule width\tableaurule height#1\hss}}}
\def\sqr{\vbox{%
  \phantomhrule\tableaustep
  \hbox{\phantomvrule\tableaustep\kern\tableaustep\phantomvrule\tableaustep}%
  \hbox{\vbox{\phantomhrule\tableauside}\kern-\tableaurule}}}
\def\squares#1{\hbox{\count0=#1\noindent\loop\sqr
  \advance\count0 by-1 \ifnum\count0>0\repeat}}
\def\tableau#1{\vcenter{\offinterlineskip
  \tableaustep=\tableauside\advance\tableaustep by-\tableaurule
  \kern\normallineskip\hbox
    {\kern\normallineskip\vbox
      {\gettableau#1 0 }%
     \kern\normallineskip\kern\tableaurule}%
  \kern\normallineskip\kern\tableaurule}}
\def\gettableau#1 {\ifnum#1=0\let\next=\null\else
  \squares{#1}\let\next=\gettableau\fi\next}
\numberwithin{equation}{section}

\newcommand{\be}{\begin{equation}}
\newcommand{\ee}{\end{equation}}

\newcommand{\IP}{\mathbb{P}}
\newcommand{\co}{{\cal O}}

\newcommand\IZ{\mathbb{Z}}
\newcommand{\IC}{\mathbb{C}}

\newcommand{\ba}{\begin{array}}
\newcommand{\ea}{\end{array}}

\newcommand{\CV}{{\mathcal V}}

\newcommand{\CK}{{\cal K}}

\newcommand{\CE}{{\cal E}} 

\newcommand{\cF}{{\cal F}}
\newcommand{\cG}{{\cal G}}
 
\newcommand{\bal}{\begin{aligned}}
\newcommand{\eal}{\end{aligned}}
\newcommand{\me}{{\mathfrak{E}}}
\newcommand{\mf}{\mathfrak{F}}

\newcommand{\mv}{{\mathfrak{V}}}
\newcommand{\mfu}{\mathfrak{U}}
\newcommand{\mfl}{\mathfrak{L}}

\newcommand{\msa}{\mathsf{a}}
\newcommand{\msb}{\mathsf{b}}
\newcommand{\msc}{\mathsf{c}}
\newcommand{\dbar}{\overline{\partial}}
\newcommand{\wCC}{{\widetilde{\mathcal{C}}}}
\newcommand{\wP}{{\widetilde{P}}}
\newcommand{\mscc}{\mathsf{c}}
\DeclareFontFamily{U}{rsf}{}
\DeclareFontShape{U}{rsf}{m}{n}{
  <5> <6> rsfs5 <7> <8> <9> rsfs7 <10-> rsfs10}{}
\DeclareMathAlphabet\Scr{U}{rsf}{m}{n}
\def\Ext{\operatorname{Ext}}
\def\Hom{\operatorname{Hom}}
\def\sExt{\operatorname{\Scr{E}\!\textit{xt}\,}}
\def\sHom{\operatorname{\Scr{H}\!\!\textit{om}}}
\begin{document}
\preprint{    {\tt hep-th/0606180}
}
\title{D-Brane Superpotentials in Calabi-Yau Orientifolds
}
\author{Duiliu-Emanuel Diaconescu, Alberto Garcia-Raboso,\\
              Robert L. Karp and Kuver Sinha}
\oneaddress{
      {\centerline{\it Department of Physics and Astronomy, Rutgers University}}
      \smallskip
      {\centerline{\it Piscataway, NJ 08854 USA}}}
\date{}

\Abstract{
We develop computational tools for the tree-level superpotential 
of \textbf{B}-branes in Calabi-Yau orientifolds. 
Our method is based on a systematic implementation of the 
orientifold projection in the geometric approach of Aspinwall and 
Katz. In the process we lay down some ground rules for orientifold 
projections in the derived category.
}

\maketitle 

\tableofcontents

\section{Introduction}\label{s:1}

D-branes in Type IIB orientifolds are an important 
ingredient in constructions of string vacua. A frequent problem 
arising in this context is the computation of the tree-level 
superpotential for holomorphic D-brane configurations. 
This is an important question for both realistic model building 
as well as  dynamical supersymmetry breaking. 

Various computational methods for the tree-level superpotential
have been proposed in the literature. A geometric approach 
which identifies the superpotential with a three-chain period 
of the holomorphic $(3,0)$-form has been  investigated in
\cite{Donaldson:1996kp,Witten:1997ep,clemens-2002-,
Kachru:2000ih,Kachru:2000an}. 
A related method, based on two-dimensional holomorphic 
Chern-Simons theory, has been developed in 
\cite{Aganagic:2000gs,Cachazo:2001gh,Cachazo:2001sg,Dijkgraaf:2002fc}. 
The tree-level superpotential for fractional brane configurations 
at toric  Calabi-Yau singularities has been computed in 
\cite{Douglas:1997de,Klebanov:1998hh,Morrison:1998cs,Greene:1998vz,
Beasley:1999uz,Feng:2000mi}. Using exceptional collections, 
one can also compute the superpotential for non-toric 
del Pezzo singularities \cite{Wijnholt:2002qz,Herzog:2003zc,
Verlinde:2005jr,Wijnholt:2005mp}. 
Perturbative disc computations
for superpotential interactions have been performed in 
\cite{Brunner:1999jq,Brunner:2000wx,Douglas:2002fr}. 
Finally, a mathematical approach based on versal 
deformations has been developed in \cite{Katz} and 
extended to matrix valued fields in \cite{Ferrari}. 

A systematic approach encompassing all these cases follows 
from the algebraic structure of \textbf{B}-branes on Calabi-Yau 
manifolds. Adopting the point of view that \textbf{B}-branes
form a triangulated differential graded category
\cite{Kontsevich,Sharpe:1999qz,Douglas:2000gi,Lazaroiu:2001jm,
Lazaroiu:2001rv,AL}
the computation of the superpotential is equivalent to the 
computation of a minimal $A_\infty$ structure for the 
D-brane category \cite{KS,Pol,Lazaroiu:2001nm,Fukaya,Herbst:2004jp}.

This approach has been employed in the Landau-Ginzburg D-brane 
category \cite{Ashok:2004xq,Hori:2004ja,Herbst:2004zm}, and 
in the derived category of coherent sheaves \cite{AK,AF}. 
These are two of  the various phases that appear
in the moduli space of a generic
${\cal N}=2$ Type II compactification.
In particular, Aspinwall and Katz \cite{AK}  developed 
a general computational approach for the superpotential, 
in which the $A_\infty$ products are computed using a 
$\check{\mathrm{C}}$ech cochain model for the off-shell open string 
fields. 

The purpose of the present paper is to apply a similar strategy 
for D-branes wrapping holomorphic curves in Type II orientifolds. 
This requires a basic understanding of the orientifold projection  
in the derived category, which is the subject of section \ref{s:2}. 
In section \ref{s:3} we propose a computational scheme for 
the superpotential in orientifold models. This relies on 
a systematic implementation of the orientifold 
projection in the calculation of the $A_\infty$ structure.

We show that the natural algebraic framework for 
deformation problems in orientifold models relies 
on $L_\infty$ rather than $A_\infty$ structures. 
This observation leads to a simple prescription for  
the D-brane superpotential in the presence of an 
orientifold projection: one has to evaluate the superpotential 
of the underlying unprojected theory on invariant on-shell 
field configurations. This is the main conceptual result of the paper,
and its proof necessitates the introduction of 
a lengthy abstract machinery.

Applying our prescription in practice requires some extra
work. The difficulty stems from the fact that while the 
orientifold action is geometric on the Calabi-Yau, it is {\em not} 
naturally geometric at the level of the derived category. Therefore, 
knowing the superpotential in the original theory does not 
trivially lead to the superpotential of the orientifolded theory. 
To illustrate this point we compute the superpotential in two 
different cases. Both will involve D-branes wrapping rational
curves, the difference will be in the way these curves are obstructed
to move in the ambient space.
 
The organization of the paper is as follows.  Section~2 reviews
the construction of the categorical framework in which we wish to
impose the orientifold projection, as well as how to do the latter.
Section~3 describes the calculation of the D-brane superpotential
in the presence of the projection. Finally, section~4 offers concrete
computations of the D-brane superpotential for obstructed curves in 
Calabi-Yau orientifolds.

{\it Acknowledgments.} We would like to thank Mike Douglas for
valuable discussions and suggestions. 
D.E.D. was partially supported by NSF grant PHY-0555374-2006.
R.L.K. was supported in part by the DOE grant DE-FG02-96ER40949.

\section{D-Brane Categories and Orientifold Projection}\label{s:2}

This section will be concerned with general aspects of 
topological \textbf{B}-branes in the presence of an orientifold 
projection. Our goal is to find a natural formulation for 
the orientifold projection in D-brane categories. 

For concreteness, we will restrict ourselves to the 
category of topological \textbf{B}-branes on a Calabi-Yau 
threefold $X$, but our techniques extend to higher dimensions.
In this case, the D-brane category is the derived category of coherent
sheaves on $X$ \cite{Kontsevich,Douglas:2000gi}. 
In fact, a systematic off-shell construction of the D-brane 
category \cite{Lazaroiu:2001jm,Lazaroiu:2001rv}
shows that the category in question is actually larger than 
the derived category. In addition to complexes, one has 
to also include twisted complexes as defined in 
\cite{BK}. We will show below that the off-shell approach  
is the most convenient starting point for a systematic  
understanding of the orientifold projection. 

\subsection{Review of D-Brane Categories}

Let us begin with a brief review of the off-shell construction
of D-brane categories 
\cite{BK,Lazaroiu:2001jm,Lazaroiu:2001rv}. It should be
noted at the offset that there are several different models 
for the D-brane category, depending on the choice of 
a fine resolution of the structure sheaf $\co_X$. 
In this section we will  
work with the Dolbeault resolution, which is closer to the 
original formulation of the  
boundary topological \textbf{B}-model \cite{EW:CS}. This model 
is very convenient for the conceptual understanding 
of the orientifold projection, but it is unsuitable for 
explicit computations. In Section~4 we will employ a 
$\check{\mathrm{C}}$ech cochain model for computational
purposes, following the path pioneered in \cite{AK}. 

Given the threefold $X$, one first defines a differential graded 
category $\CC$ as follows 
\[
\begin{aligned} 
& \mathrm{Ob}(\CC) \colon \textrm{holomorphic  vector  bundles }
(E,\dbar_E) \textrm{ on } X \\
& \mathrm{Mor}_\CC\left((E,\dbar_E),(F,\dbar_F)\right)  
=\ \left(\oplus_p\, A^{0,p}_X(\sHom_X(E,F)), \ \dbar_{EF}\right)
\end{aligned}
\]
where we have denoted by $\dbar_{EF}$ the induced Dolbeault operator 
on $\sHom_X(E,F)$-valued $(0,p)$ forms.\footnote{
$\sHom_X(E,F)$ is the sheaf Hom of $E$ and $F$, viewed as sheaves.}
The space of morphisms is a $\IZ$-graded differential complex. In order to
simplify the notation we will denote the objects of $\CC$ by $E$, the data of
an integrable Dolbeault operator $\dbar_E$ being implicitly understood. 

The composition of morphisms in $\CC$ is defined by exterior multiplication 
of bundle valued differential forms. For any object $E$ composition 
of morphisms determines an associative algebra structure on 
the endomorphism space $\mathrm{Mor}_\CC(E,E)$. This product is 
compatible with the differential, therefore we obtain a differential 
graded associative algebra structure (DGA) on $\mathrm{Mor}_\CC(E,E)$.
 
At the next step, we construct the \emph{shift completion} $\wCC$ 
of $\CC$, which is a category of holomorphic vector bundles on $X$ 
equipped with an integral grading. 
\[
\begin{aligned}
& \mathrm{Ob}(\wCC) \colon \textrm{pairs } (E,n), \textrm{with $E$ an object of }
\CC \textrm{ and } n\in \IZ \\
& \mathrm{Mor}_{\wCC}((E,n),(F,m)) = \mathrm{Mor}_\CC(E,F)[n-m].
\end{aligned}
\] 
The integer $n$ is the boundary ghost number introduced in 
\cite{Douglas:2000gi}. Note that for a homogeneous element 
\[
f \in \mathrm{Mor}^k_\wCC((E,n),(F,m))
\]
we have
\[
k = p +(m-n) 
\] 
where $p$ is the differential form degree of $f$. 
The degree $k$ represents the total ghost number of 
the field $f$ with respect to the bulk-boundary BRST operator.
In the following we will use the notations 
\[ 
|f| = k, \qquad c(f) = p, \qquad h(f) = m-n.
\]

The composition of morphisms in $\wCC$ differs from the
composition of morphisms in $\CC$ by a sign, which will
play an important role in our construction. Given two
homogeneous elements 
\[ 
f\in \mathrm{Mor}_\wCC((E,n),(E',n'))\qquad 
g \in \mathrm{Mor}_\wCC((E',n'),(E'',n''))
\]
one defines the composition
\be\label{eq:signruleA}
(g\comp f)_\wCC = (-1)^{h(g)c(f)} (g\comp f)_\CC \,.
\ee
This choice of sign leads to the graded Leibniz rule
\[ 
\dbar_{EE''}(g\comp f)_\wCC = \left(\dbar_{E'E''}(g)\comp f\right)_\wCC 
+(-1)^{h(g)}\left (g\comp \, \dbar_{EE'}(f)\right)_\wCC\,.
\]

Now we construct a pre-triangulated DG category $\textrm{Pre-Tr}(\wCC)$
of twisted complexes as follows 
\[ 
\begin{aligned} 
& \mathrm{Ob}\left(\textrm{Pre-Tr}(\wCC)\right) \colon \quad
\mbox{
\begin{minipage}[0pt]{4in}
finite collections of the form\\
$\left\{(E_i,n_i,q_{ji})|\ q_{ji}\in  \mathrm{Mor}^1_{\wCC}((E_i,n_i),(E_j,n_j))\right\}$\\
where the $q_{ji}$ satisfy the Maurer-Cartan equation\\
$\dbar_{E_iE_j} (q_{ji}) + \sum_{k} (q_{jk}\comp q_{ki})_{\wCC} =0$.
\end{minipage}
} \\
&\mathrm{Mor}_{\textrm{Pre-Tr}(\wCC)}((E_i,n_i,q_{ji}), (F_i,m_i,r_{ji})) 
=\left(\bigoplus_{i,j} \mathrm{Mor}_{\wCC}((E_i,n_i),(F_j,m_j)), \ Q\right)
\end{aligned}
\]
where the differential $Q$ is defined as
\[
Q(f) = \dbar_{E_iF_j}(f) + \sum_{k} (r_{kj} \comp f)_{\wCC} - 
(-1)^{|f|} (f \comp q_{ik})_{\wCC},\qquad f \in 
\mathrm{Mor}_{\wCC}((E_i,n_i),(F_j,m_j)).
\]
$|f|$ is the degree of $f$ in $\mathrm{Mor}_{\wCC}((E_i,n_i),(F_j,m_j))$
from above. For each object, the index $i$ takes finitely many values
between $0$ and some maximal value which depends on the object.
Note that $Q^2=0$ because $\{q_{ji}\},\{r_{ji}\}$ satisfy the Maurer-Cartan
equation. Composition of morphisms in $\textrm{Pre-Tr}(\wCC)$ reduces to
composition of morphisms in $\wCC$.

Finally, the triangulated D-brane category $\mathcal{D}$ has by definition 
the same objects as $\textrm{Pre-Tr}(\wCC)$, while its morphisms are given
by the zeroth cohomology under $Q$ of the morphisms of $\textrm{Pre-Tr}(\wCC)$:
\be\label{eq:triangulatedA}
\begin{aligned} 
& \mathrm{Ob}\left(\mathcal{D}\right) = \mathrm{Ob}\left(\textrm{Pre-Tr}(\wCC)\right) \\
& \mathrm{Mor}_{\mathcal{D}}((E_i,n_i,q_{ji}),(F_i,m_i,r_{ji})) =
H^0\left(Q,\, \mathrm{Mor}_{\textrm{Pre-Tr}(\wCC)}((E_i,n_i,q_{ji}),(F_i,m_i,r_{ji}))
\right).
\end{aligned}
\ee

The bounded derived category of coherent sheaves
$D^b(X)$ is a full subcategory of ${\mathcal D}$. To see this consider 
the objects of the form 
$(E_i,n_i,q_{ji})$ such that 
\be\label{eq:subcat}
n_i = -i, \qquad q_{ji} \neq 0 \ \Leftrightarrow \ j=i-1. 
\ee
Since $q_{ji}\in \mathrm{Mor}^1_\wCC((E_i,n_i),
(E_j,n_j))$, the second condition in \eqref{eq:subcat} 
implies that their differential form degree must be $0$. 
The Maurer-Cartan equation for such objects reduces to 
\[
\dbar_{E_{i}E_{i-1}} q_{i-1,i} =0, \qquad 
(q_{i-1,i}\comp q_{i,i+1})_\wCC =0. 
\]
Therefore the twisted complex $(E_i,n_i,q_{ji})$ is in fact a 
complex of holomorphic vector bundles 
\be\label{eq:complexA}
\xymatrix@C=8mm@M=1mm@1{
& \cdots \ar[r]&  E_{i+1} \ar[rr]^{q_{i,i+1}} &&  
E_{i}\ar[rr]^{ q_{i-1,i}} && E_{i-1} \ar[r]
& \cdots\\}
\ee
We will use the alternative notation 
\be\label{eq:complexB} 
\xymatrix@=10mm@1{
& \cdots \ar[r]&  E_{i+1} \ar[r]^-{d_{i+1}} & 
E_{i}\ar[r]^-{d_i} &  E_{i-1} \ar[r]
& \cdots\\}
\ee
for complexes of vector bundles, and also denote them by the corresponding Gothic letter, here $\me$.

One can easily check that the morphism space \eqref{eq:triangulatedA} 
between two twisted complexes
of the form \eqref{eq:subcat} reduces to the hypercohomology group  
of the local Hom complex $\sHom({\mathfrak E},{\mathfrak F})$
\be\label{eq:triangulatedB} 
\mathrm{Mor}_{\mathcal{D}}((E_i,n_i,q_{ji}),(F_i,m_i,r_{ji})) 
\simeq\mathbb{H}^0(X, \, \sHom({\mathfrak E},{\mathfrak F})).
\ee
As explained in \cite{AL}, this hypercohomology group is isomorphic 
to the derived morphism space $\Hom_{D^b(X)}(\me,\mf)$.
Assuming that $X$ is smooth and projective, any derived object has a 
locally free resolution, hence $D^b(X)$ is a full
subcategory of $\mathcal{D}$. 

\subsection{Orientifold Projection}

Now we consider orientifold projections from the D-brane category
point of view. A similar discussion of orientifold projections 
in matrix factorization categories has been outlined in \cite{Hori}. 

Consider a four dimensional $N=1$ IIB orientifold obtained 
from an $N=2$ Calabi-Yau compactification by gauging 
a discrete symmetry of the form $$(-1)^{\epsilon F_L}\Omega\,\sigma$$
with $\epsilon =0,1$.
Employing common notation, $\Omega$ denotes world-sheet parity,
$F_L$ is the left-moving fermion number and $\sigma\colon X \to X$ is
a holomorphic involution of $X$ satisfying 
\be\label{eq:orproj}
\sigma^* \Omega_X = (-1)^\epsilon\, \Omega_X,
\ee
where $\Omega_X$ is the holomorphic $(3,0)$-form of the Calabi-Yau. 
Depending on the value of $\epsilon$, there are two classes of models
to consider \cite{Grimm:2004uq}:
\begin{enumerate}
\item $\epsilon=0$: theories with $O5/O9$ orientifolds planes, in which
the fixed point set of $\sigma$ is either one or three complex
dimensional;
\item $\epsilon=1$: theories with $O3/O7$ planes, with $\sigma$ leaving
invariant zero or two complex dimensional submanifolds of $X$.
\end{enumerate}

Following the same logical steps as in the previous subsection, 
we should first find the action of the orientifold projection  
on the category $\CC$, which is the starting point of the 
construction. The action of parity on the K-theory class of a D-brane  
has been determined in \cite{EW:Ktheory}. The world-sheet parity
$\Omega$ maps $E$ to the dual vector bundle $E^\roof$.   
If $\Omega$ acts simultaneously with a holomorphic involution
$\sigma\colon X\to X$, the bundle $E$ will be mapped to
$\sigma^\ast(E^\roof)$. If the projection also involves a $(-1)^{F_L}$
factor, a brane with Chan-Paton bundle $E$ should be mapped to
an anti-brane with Chan-Paton bundle $P(E)$. 

Based on this data, we define the action of parity on $\CC$ to be 
\be\label{eq:paractA}
\begin{aligned} 
P\colon  E & \mapsto P(E) = \sigma^\ast(E^\roof) \\
P\colon  f \in \mathrm{Mor}_{\CC}(E,F) & \mapsto 
\sigma^\ast(f^\roof) \in \mathrm{Mor}_{\CC}(P(F),P(E))\\
\end{aligned}
\ee
It is immediate that $P$ satisfies the following 
compatibility condition with respect to 
composition of morphisms in $\CC$:
\be\label{eq:paractB} 
P((g\comp f)_\CC) = (-1)^{c(f)c(g)}\left(P(f)\comp P(g)\right)_\CC
\ee
for any homogeneous elements $f$ and $g$. 
It is also easy to check that $P$ preserves the differential 
graded structure, i.e., 
\be\label{eq:dgfctA}
P(\dbar_{EF}(f)) = \dbar_{P(F)P(E)}(P(f)). 
\ee
Equation \eqref{eq:paractB} shows that $P$ is not
a functor in the usual sense. Since it is compatible with 
the differential graded structure, it should be interpreted as a 
functor of $A_\infty$ categories \cite{Keller}. Note however that $P$ is 
``almost a functor'': it fails to satisfy the compatibility 
condition with composition of morphisms only by a sign. 
For future reference, 
we will refer to $A_\infty$ functors satisfying a graded compatibility 
condition of the form \eqref{eq:paractB} as \emph{graded functors}.

The category $\CC$ does not contain enough
information to make a distinction between branes and anti-branes. 
In order to make this distinction, we have to assign each 
bundle a grading, that is we have to work in the category  
$\wCC$ rather than $\CC$. By convention, the objects $(E,n)$ with $n$ even are
called branes, 
while those with $n$ odd are called anti-branes. 

We will take the action of the orientifold projection on the objects of $\wCC$ 
to be   
\be\label{eq:paractC}
\wP\colon (E,n)\mapsto (P(E),m-n)
\ee 
where we have introduced an integer shift $m$ which is correlated with 
$\epsilon$ from \eqref{eq:orproj}:
\be\label{eq:corr}
m \equiv \epsilon \ \mathrm{mod} \ 2.
\ee
This allows us to treat both cases 
$\epsilon=0$ and $\epsilon=1$ in a unified framework. 

We define the action of $P$ on a morphisms 
$f\in \mathrm{Mor}_\wCC((E,n),(E',n'))$
as the following graded dual:
\be\label{eq:paractD} 
\wP(f)= -(-1)^{n'h(f)} P(f),
\ee
where $P(f)$ was defined in \eqref{eq:paractA}.\footnote{
There is no a priori justification for the particular
sign we chose, but as we will see shortly, it leads 
to a graded functor. 
A naive generalization of \eqref{eq:paractA} ignoring this 
sign would not yield a graded functor. }
Note that the graded dual has been used in a similar context in 
\cite{Hori}, where the orientifold projection is implemented 
in matrix factorization categories. 

With this definition, we have the following:
\begin{prop}
$\wP$ is a graded functor on 
$\wCC$ satisfying
\be\label{eq:paractE} 
\wP((g\comp f)_\wCC) = -(-1)^{|f||g|} (\wP(f) \comp \wP(g))_\wCC
\ee
for any homogeneous elements 
\[
f\in \mathrm{Mor}_{\wCC}((E,n),(E',n')),\qquad
g \in \mathrm{Mor}_{\wCC}((E',n'),(E'',n'')).
\]
\end{prop}
\begin{proof}
It is clear that $\wP$ is compatible with the differential graded structure 
of $\wCC$ since the latter is inherited from $\CC$. 

Next we prove \eqref{eq:paractE}. First we have:
\[
\begin{aligned}
\wP((g\comp f)_\wCC) & = -(-1)^{n'' h(g\comp f)} 
P((g\comp f)_\wCC)\qquad 
\mbox{\footnotesize by \eqref{eq:paractD}}
\cr
& = -(-1)^{n'' h(g\comp f) + h(g)c(f)}
 P(  (g\comp f)_\CC)
\qquad \mbox{\footnotesize by \eqref{eq:signruleA}} 
\cr
& = -(-1)^{n'' h(g\comp f) + h(g)c(f) + c(f)c(g)}
 (P(f)\comp P(g))_\CC
\qquad \mbox{\footnotesize by  \eqref{eq:paractB}}
\end{aligned}
\]
On the other hand
\[
\begin{aligned}
(\wP(f)\comp \wP(g))_\wCC & = 
(-1)^{n'h(f)+n''h(g)} (P(f)\comp P(g))_\wCC
\qquad \mbox{\footnotesize by \eqref{eq:paractD}}
\cr
& = (-1)^{n'h(f)+n''h(g)} (-1)^{ h(P(f))c(P(g))}
(P(f)\comp P(g))_\CC
\qquad \mbox{\footnotesize by \eqref{eq:signruleA}} 
\end{aligned}
\]
But 
$$h(g\comp f)=h(f)+h(g),\quad h(P(f))=h(f),\quad c(P(g))=c(g).$$
Now \eqref{eq:paractE} follows from 
\[
n''(h(f)+h(g)) - n'h(f)-n''h(g) = (n''-n')h(f) = h(g)h(f)
\]
and 
\[
|f||g| = (h(f)+c(f))(h(g)+c(g)).
\]
\end{proof}

The next step is to determine the action of $P$
on the pre-triangulated category $\mbox{Pre-Tr}(\wCC)$.
We denote  this action by $\widehat{P}$.
The action of $\widehat{P}$ on objects is defined simply by 
\be\label{eq:paractF}
\begin{aligned}
(E_i,n_i,q_{ji}) \mapsto (P(E_i), m-n_i,\wP(q_{ji}))
\end{aligned}
\ee 
Using equations \eqref{eq:dgfctA} and \eqref{eq:paractE}, 
it is straightforward to show that the action of $\widehat{P}$ 
preserves the Maurer-Cartan equation, that is 
\[
\dbar_{E_iE_j} (q_{ji}) + \sum_{k} (q_{jk}\comp q_{ki})_{\wCC} =0 \
\Rightarrow \ \dbar_{P(E_j) P(E_i)} \wP(q_{ji}) + 
\sum_k (\wP(q_{ki})\comp \wP(q_{jk}))_{\wCC}=0,
\]
since all $q_{ji}$ have total degree one.
Therefore this transformation is well defined on objects. 
The action on morphisms is also straightforward
\begin{multline}\label{eq:paractG}
f \in  \oplus_{i,j} \mathrm{Mor}_{\wCC}((E_i,n_i),(F_j,m_j)) \\
\mapsto \widehat{P}(f)=\wP(f) \in \oplus_{i,j} 
\mathrm{Mor}_{\wCC}((P(F_j),m-m_j),(P(E_i),m-n_i)).\qquad
\end{multline}
Again, equations \eqref{eq:dgfctA}, \eqref{eq:paractE} 
imply that this action preserves the differential 
\[
\qquad Q(f) = \dbar_{E_iF_j}(f) + \sum_{k} (r_{kj} \comp f)_{\wCC} - 
(-1)^{|f|} (f \comp q_{ik})_{\wCC}
\]
since $\{q_{ji}\}$, $\{r_{ji}\}$ have degree one. This means we have 
\be\label{eq:paractH} 
\widehat{P}(Q(f)) = \dbar_{P(F_j) P(E_i)} \wP(f) + \sum_{k} 
(\wP(q_{ik})\comp \wP(f))_\wCC - (-1)^{|\wP(f)|} (\wP(f)\comp 
\wP(r_{kj}) )_\wCC
\ee

For future reference, let us record some 
explicit formulas for complexes of vector bundles. 
A complex 
\[\me\colon
\xymatrix@=10mm@1{
& \cdots \ar[r]&  E_{i+1} \ar[r]^-{d_{i+1}} & 
E_{i}\ar[r]^-{d_i} &  E_{i-1} \ar[r]
& \cdots\\}
\]
in which $E_i$ has degree $-i$ is mapped to the complex
\be\label{eq:paractI}
\widehat{P}(\me)\colon
\xymatrix{
& \cdots \ar[r]&  P(E_{i-1}) \ar[rr]^{\wP(d_i)} & &
P(E_i)\ar[rr]^{ \wP(d_{i+1})} & & 
P(E_{i+1}) \ar[r]
& \cdots\\}
\ee
where $\wP(d_i)$ is determined by (\ref{eq:paractD})
\[
\wP(d_i) = (-1)^{i}\,\sigma^*(d_i^\roof) 
\] 
and $P(E_i)$ has degree $i-m$. 
Applying $\widehat{P}$ twice yields the complex 
\be\label{eq:paractJ} 
\widehat{P}^2(\me)\colon
\xymatrix{
& \cdots \ar[r]&  E_{i+1} 
\ar[rr]^{\wP^2(d_{i+1})} &&
E_{i}\ar[rr]^{ \wP^2(d_{i})} &&   E_{i-1} \ar[r]
& \cdots\\}
\ee
where 
\[
\begin{aligned}
\wP^2(d_{i}) 
& = (-1)^{m+1} d_i\,. 
\end{aligned}
\]
Therefore $\widehat{P}^2$ is not equal to the identity functor, but there 
is an isomorphism of complexes 
$J\colon \widehat{P}^2(\me) \to \me :$
\be\label{eq:paractK} 
\xymatrix{ 
 & \cdots \ar[r]& E_{i+1}\ar[d]^{{J_{i+1}}} 
\ar[rrr]^{\wP^2(d_{i+1})} &&& 
E_{i}\ar[rrr]^{ \wP^2(d_i)}\ar[d]^{J_i} 
&&&   E_{i-1} \ar[d]^{J_{i-1}}\ar[r]
& \cdots\\
& \cdots \ar[r]&  E_{i+1} \ar[rrr]^{d_{i+1}} &&& 
E_{i}\ar[rrr]^{ d_i} &&&   E_{i-1} \ar[r]
& \cdots\\}
\ee
where
\[
J_i = (-1)^{(m+1)i}\chi\,\mathrm{Id}_{E_i}. 
\]
and $\chi$ is a constant. Notice that $J^{-1}\colon \me \to \widehat{P}^2(\me)$,
and that $\widehat{P}^4 = \mathrm{Id}_{D^b(X)}$ implies that  also
$J\colon \widehat{P}^2(\widehat{P}^2(\me)) = \me \to \widehat{P}^2(\me)$. Requiring both to be equal
constrains $\chi$ to be $(-1)^\omega$ with $\omega = 0, 1$.
This sign cannot be fixed using purely algebraic considerations,
and we will show in section \ref{s:4} how it encodes the difference
between $SO/Sp$ projections. 
In functorial language, this means that there is an isomorphism of functors 
$J: \widehat{P}^2 \to \mathrm{Id}_{D^b(X)}$.
 
We conclude this section with a brief summary of the above 
discussion, and a short  remark on possible generalizations.
To simplify notation, in the rest of the paper we drop the decorations of
the various $P$'s. In other words both $\widehat{P}$ and $\wP$ will be
denoted by $P$. Which $P$ is meant will always be clear from the context. 

\begin{enumerate} 
\item{} The orientifold projection in the derived 
category is a graded contravariant 
functor $P\colon D^b(X) \to D^b(X)^{op}$ which acts on locally free 
complexes as in equation \eqref{eq:paractI}. Note that this transformation 
is closely related to the derived functor 
\[
\mathbf{L}\sigma^* \comp \mathbf{R}\!\sHom(- , \co_X)[m]. 
\]
The difference resides in the alternating signs $(-1)^i$ in the 
action of $P$ on differentials, according to \eqref{eq:paractI}. 
From now on we will refer to $P$ as a graded derived functor. 

\item{} There is an obvious generalization of this construction which 
has potential physical applications. One can further compose $P$
with an auto-equivalence $\CA$ of the derived category so that the 
resulting graded functor $P\comp \CA$ has its square isomorphic to the identity. 
This would yield a new class of orientifold models, possibly without 
a direct geometric interpretation. The physical implications of 
this construction will be explored in a separate publication. 
\end{enumerate}

\noindent
In the remaining part of this section we will consider the case 
of D5-branes wrapping holomorphic curves in more detail. 

\subsection{$O5$ models}\label{s:2.3}

In this case we consider holomorphic involutions 
$\sigma\colon X\to X$ whose fixed point set consists 
of a finite collection of holomorphic curves in $X$. We will be 
interested in D5-brane configurations supported on a smooth
component $C\simeq \IP^1$ of the fixed locus, that are preserved
by the orientifold projection.
We describe such a configuration by a one term complex 
\be\label{eq:dfiveA}{  {i_* V}}\ee
where $V\to C$ is the Chan-Paton vector 
bundle on $C$, and 
$i\colon C\hookrightarrow X$ is the embedding of $C$ into $X$. 

Since $C\simeq \IP^1$, by Grothendieck's theorem any holomorphic bundle $V$ decomposes in a 
direct sum of line bundles. Therefore, for the time being, we take  
\be\label{eq:CPbundleA}
V \simeq \co_C(a)
\ee
for some $a\in \IZ$. 
We will also make the simplifying assumption 
that $V$ is the restriction of a bundle $V'$
on $X$ to $C$, i.e.,
\begin{equation}\label{ek1}
V=i^* V'.
\end{equation}
This is easily satisfied if $X$ is a complete intersection 
in a toric variety $Z$, in which case $V$ can be chosen to be the restriction 
of bundle on $Z$. 

In order to write down the parity action on this D5-brane
configuration, we need a locally free resolution $\me$ for $i_* V=i_* \co_C(a)$. Let 
\be\label{eq:resolA}
\xymatrix{
\mv: & 0 \ar[r] & \CV_n \ar[r]^-{d_n} & \CV_{n-1} \ar[r]^-{d_{n-1}} &
\cdots \ar[r]^-{d_2} & \CV_1 \ar[r]^-{d_1} & \poso{\CV_0} \ar[r] &  0}
\ee
be a locally free resolution of $i_*\co_C$.\footnote{
We usually underlined the $0$th position in a complex.}
The degree of the term 
$\CV_k$ to be $(-k)$, for $k=0,\ldots,n$. 
Then the complex $\me$
\be\label{eq:resolB} 
\xymatrix{
\me:& 0\ar[r]&  \CV_n(a) \ar[r]^-{d_n} & \CV_{n-1}(a)\ar[r]^-{d_{n-1}} &  
\cdots\ar[r]^-{d_2} & \CV_1(a) \ar[r]^-{d_1}
& \poso{\CV_0(a)} \ar[r] &  0\\}
\ee
is a locally free resolution of $i_* \co_C(a)$. 

The image of
\eqref{eq:resolB} under the orientifold projection  
is the complex $\CP(\me)$:
\be\label{eq:pdfiveA}\nonumber
\xymatrix@C=12mm@M=1mm{
0\ar[r]& { {\sigma^*\CV_0^\roof(-a)}}\ar[r]^{-\sigma^* d_1^\roof} &
*{\;\sigma^*\CV_{1}^\roof(-a)} \ar[r]^(0.6){\sigma^* d_2^\roof}  & \cdots }
\ee
\be
\xymatrix@C=21mm@M=1mm{
\cdots\ar[r]^(0.35){(-1)^{n-1}\sigma^*d_{n-1}^\roof} &
*{\;\sigma^*\CV_{n-1}^\roof(-a)} \ar[r]^{(-1)^n\sigma^* d_n^\roof}
&  \sigma^*\CV_n^\roof(-a) \ar[r] &  0\\}
\ee
The term $\sigma^*{\CV_k^\roof}(-a)$ has degree $k-m$.

\begin{lemma}\label{c:1}\footnote{We give an alternative derivation of this
result in Appendix~\ref{a:1}. That proof is very abstract, and hides all the
details behind the powerful machinery of Grothendieck duality. On the other
hand, we will be using the details of this lengthier derivation in our 
explicit computations in Section~\ref{s:4}.}
The complex \eqref{eq:pdfiveA} is quasi-isomorphic to 
\be\label{eq:twistdualA}
{  {i_* \left(V^\roof \otimes K_C\right)}}[m-2],
\ee
where $K_C \simeq \co_C(-2)$ is the canonical bundle of $C$. 
\end{lemma}
\begin{proof}
As noted below \eqref{eq:paractI}, \eqref{eq:pdfiveA} is isomorphic 
to $\sigma^*(\me^\roof)[m]$. Since $C$ is 
pointwise fixed by $\sigma$, it suffices to show that the dual of the 
locally free resolution \eqref{eq:resolA} is quasi-isomorphic to 
${  {i_*K_C}}[-2]$. The claim then follows from the adjunction formula:
\begin{equation}
i_* V=i_*(V\otimes \co_C)=i_*(i^*V' \otimes \co_C)=V' \otimes i_*\co_C
\end{equation}
and the simple fact that 
$i^* (V'^\roof )=V^\roof $.

Let us compute  $(i_*\co_C)^\roof$ using the locally free resolution
\eqref{eq:resolA}. 
The cohomology in degree $k$ of the complex 
\be\label{eq:dualres}
\mv^\roof\colon\quad 0\to ({\CV_0})^\roof 
\to {(\CV_1)}^\roof \to \cdots \to {(\CV_{n})}^\roof \to 0
\ee
is isomorphic to the local Ext sheaves ${\sExt}_X^k(i_*\co_C,\co_X)$. 
According to \cite[Chapter 5.3, pg 690]{GH} 
these are trivial except for $k=2$, in which case 
\[ 
{\sExt}_X^2(\co_C,\co_X) \simeq i_*\CL,
\]
for some  line bundle $\CL$ on $C$. 

To determine $\CL$, it suffices to 
compute its degree on $C$, which is an easy application of the 
Grothendieck-Riemann-Roch theorem. We have 
\[
i_!(\mathrm{ch}(\CL)\mathrm{Td}(C))= \mathrm{ch}(i_*\CL) \mathrm{Td}(X) .
\]
On the other hand, by construction
\[
\mathrm{ch}_{m}(i_*\CL) = \mathrm{ch}_m(\mv^\roof) = (-1)^m
\mathrm{ch}_{m}(\mv) = (-1)^m\mathrm{ch}_m(i_*\co_C).
\]
Using these two equations, we find 
\[
\mathrm{deg}(\CL)=-2 \ \Rightarrow \CL\simeq K_C. 
\]
This shows that $\mv^\roof$ 
has nontrivial cohomology $i_*K_C$ only in degree $2$. 

Now we establish that the complex \eqref{eq:dualres} is quasi-isomorphic to 
${  {i_*K_C}}[-2]$, by constructing such a map of complexes. Consider the 
restriction of the complex \eqref{eq:dualres}
to $C$. Since all terms are locally free, we obtain a complex 
of holomorphic bundles on $C$ whose cohomology is isomorphic 
to $K_C$ in degree $2$ and trivial in all other degrees. 
Note that the kernel $\CK$ of the map 
\[ 
{\CV_2}^\roof|_C \to {\CV_3}^\roof|_C
\] 
is a torsion free sheaf on $C$, therefore it must be locally free. 
Hence $\CK$ is a sub-bundle of ${\CV_2}^\roof|_C$. Since $C\simeq \IP^1$, 
by Grothendieck's theorem
both ${\CV_2}^\roof|_C$ and $\CK$ are isomorphic to direct sums 
of line bundles. This implies that $\CK$ is in fact a direct summand 
of ${\CV_2}^\roof|_C$. In particular there is a surjective map 
\[ 
\rho\colon {\CV_2}^\roof|_C \to \CK.
\]
Since $H^2(\mv^\roof|_C)= K_C$ we also have a surjective map 
$\tau\colon\CK \to K_C$. By construction then 
$\tau\comp\rho\colon\mv^\roof|_C \to {  K_C}[-2]$ is a quasi-isomorphism.
Extending this quasi-isomorphism by zero outside $C$, we obtain a
quasi-isomorphism 
$\mv^\roof \to {  {i_*K_C}}[-2]$, which proves the lemma. 
\end{proof}

Let us now discuss parity invariant D-brane configurations. 
Given the parity action \eqref{eq:twistdualA}  
one can obviously construct such configurations by 
taking direct sums of the form 
\be\label{eq:invconfA} 
{  {i_*V}}\oplus {  {i_*(V^\roof\otimes K_C)}}[m-2]  
\ee 
with $V$ an arbitrary Chan-Paton bundle. Note that in this case we have 
two stacks of D5-branes in the covering space which are interchanged 
under the orientifold projection. 

However, on physical grounds we should also be able to construct a single 
stack of D5-branes wrapping $C$ which is preserved by the orientifold action. 
This is possible only if 
\be\label{eq:invconfB} 
m=2 \quad \mathrm{and} \quad V \simeq V^\roof \otimes K_C.
\ee
The first condition in \eqref{eq:invconfB} 
fixes the value of $m$ for this class of models. 
The second condition constrains the Chan-Paton bundle $V$ to 
\[   
V = \co_C(-1). 
\] 

Let us now consider rank $N$ Chan-Paton bundles $V$. We will focus on 
invariant D5-brane configurations given by 
\[
V = \co_C(-1)^{\oplus N}.  
\]
In this case the orientifold image $P(i_*V)= i_*(V^\roof \otimes K_C)$ is
isomorphic to $i_*V$, and the choice of an isomorphism corresponds to the choice of
a section 
\be\label{eq:section}
M \in \Hom_C(V,V^\roof\otimes K_C) \simeq \CM_N(\IC).
\ee
where $\CM_N(\IC)$ is the space of $N\times N$ complex matrices.
We have 
\[
\begin{aligned}
\Hom_C(V,V^\roof \otimes K_C) & \simeq H^0(C, S^2(V^\roof) \otimes K_C) 
\oplus H^0(C,\Lambda^2(V^\roof) \otimes K_C)\cr
& \simeq \CM_N^+(\IC) \oplus \CM^-_N(\IC) \cr
\end{aligned}
\]
where $\CM_N^{\pm}(\IC)$ denotes the space of symmetric and antisymmetric 
$N\times N$ matrices respectively. The choice of this isomorphism (up to 
conjugation) encodes the difference between $SO$ and $Sp$ projections. 
For any value of $N$ we can choose the isomorphism to be 
\be\label{eq:soproj}
M=\mathrm{I}_N \in \CM_N^+(\IC),
\ee
obtaining $SO(N)$ gauge group. 
If $N$ is even, we also have the option of choosing the antisymmetric 
matrix 
\be\label{eq:spproj}
M=i\left[\begin{array}{cc} 0 & \mathrm{I}_{N/2} \cr -\mathrm{I}_{N/2} & 0 \cr
\end{array} \right] \, \in \, \CM_N^-(\IC) 
\ee
obtaining $Sp(N/2)$ gauge group. This is a slightly more abstract
reformulation of \cite{Gimon:1996rq}. We will explain how the $SO/Sp$ 
projections are encoded in the derived formalism in sections~\ref{s:3}
and~\ref{s:4}.

\subsection{$O3/O7$ Models} 

In this case we have $\epsilon=1$, and the fixed 
point set of the holomorphic involution can have both zero and two dimensional
components. We will
consider the magnetized D5-brane configurations introduced in
\cite{DGS:landscape}. 
Suppose \[i\colon C\hookrightarrow X \qquad i':C'\hookrightarrow X\] 
is a pair of smooth rational curves mapped isomorphically into 
each other by the holomorphic involution.  
The brane configuration consists of a 
stack of D5-branes wrapping $C$, which is related by the orientifold 
projection to a stack of anti-D5-branes wrapping $C'$. We describe the stack
of D5-branes wrapping $C$ by a one term 
complex $i_*V$, with $V$ a bundle on $C$. 

In order to find the action of the orientifold group on the stack of D5-branes
wrapping $C$ we pick a locally free resolution $\me$ for $i_* V$. 
Once again  the orientifold image is obtained by applying 
the graded derived functor $P$ to $\me$. 

Applying Prop.~\ref{pa1}, we have  
\begin{lemma}\label{c:10}
$P(\me)$ is quasi-isomorphic to the one term complex
\be\label{eq:twistdualBS}
{  {i'_*(\sigma^*(V^\roof)\otimes K_{C'})}}[m-2]. 
\ee 
\end{lemma} 
\noindent 
It follows that a D5-brane configuration preserved by the orientifold projection 
is a direct sum
\be\label{eq:magbranes}
{  {i_*V}}\oplus 
{  {i'_*(\sigma^*(V^\roof)\otimes K_{C'})}}[m-2].
\ee

The value of $m$ can be determined from physical arguments by 
analogy with the previous case. We have to impose the condition 
that the orientifold projection preserves a D3-brane supported 
on a fixed point $p\in X$ as well as a D7-brane supported on 
a pointwise fixed surface $S\subset X$. 

A D3-brane supported at $p\in X$ is described by a one-term 
complex ${  {\co_{p,X}}}$, 
where $\co_{p,X}$ is a skyscraper sheaf supported at $p$.  
Again, using Prop.~\ref{pa1} one shows that 
$P(\mv)$ is quasi-isomorphic to $\co_{p,X}[m-3]$.
Therefore, the D3-brane is preserved if and only if $m=3$. 

If the model also includes a codimension 1 pointwise-fixed locus $S\subset X$, then
we have an extra condition. Let $V$ be the Chan-Paton bundle on $S$. We
describe the  invariant D7-brane wrapping 
$S$ by $\mfl \simeq {  {i_*(V)}}[k]$ for some integer $k$, where $i\colon S\to
X$ is the embedding.

Since $S$ is codimension 1 in $X$, Prop.~\ref{pa1} tells us that
\begin{equation}
P(\mfl) \simeq {  {i_*(V^\roof\otimes K_S)}}[m-k-1].
\end{equation}
Therefore invariance under $P$ requires 
\be\label{eq:invconfC}
2k=m-1\qquad V\otimes V \simeq K_S.
\ee
Since we have found $m=3$ above, it follows that  $k=1$. Furthermore, 
$V$ has to be a square root of $K_S$. 
In particular, this implies that $K_S$ must be even, or, in other words 
that $S$ must be spin. This is in agreement with the Freed-Witten 
anomaly cancellation condition \cite{FW}. 
If $S$ is not spin, one has to turn on a half integral $B$-field 
in order to cancel anomalies.  

Returning to the magnetized D5-brane configuration, 
note that an interesting situation from the 
physical point of view is the case when the curves $C$ and $C'$ coincide. 
Then $C$ is preserved 
by the holomorphic involution, but not pointwise fixed as in the previous 
subsection. We will discuss examples of such configurations in 
section~\ref{s:4}. 
In the next section we will focus on general aspects of the 
superpotential in orientifold models. 

\section{The Superpotential}\label{s:3}

The framework of D-brane categories offers a systematic 
approach to the computation of the tree-level superpotential. 
In the absence of the orientifold projection, the tree-level D-brane 
superpotential is encoded in the $A_\infty$ structure of the D-brane  
category \cite{KS,Pol,Lazaroiu:2001nm,Herbst:2004jp}. 

Given an object of the D-brane category ${\mathcal D}$, 
the space of off-shell open string states is
its space of endomorphisms in the pre-triangulated category 
$\textrm{Pre-Tr}(\wCC)$. This carries the structure of a 
$\IZ$-graded differential cochain complex. In this section we 
will continue to work with Dolbeault cochains, and also
specialize our discussion to locally free complexes $\me$
of the form  \eqref{eq:complexB}.
Then the space of off-shell open string states is given by 
\[
\mathrm{Mor}_{\textrm{Pre-Tr}(\wCC)}(\me,\me) = \oplus_{p} 
A^{0,p}({\sHom}_X(\me,\me))
\]
where \[
{\sHom}^q_X(\me,\me) = \oplus_{i}\sHom_X(E_i,E_{i-q}).
\]

Composition of morphisms defines a natural superalgebra structure 
on this endomorphism space \cite{Diaconescu:2001ze}, and
the differential $Q$ satisfies the graded Leibniz rule. 
We will denote the resulting DGA by $\CC(\me,\me)$. 

The computation of the superpotential 
is equivalent to the construction of an $A_\infty$ minimal 
model for the DGA $\CC(\me,\me)$. Since this formalism 
has been explained in detail in the physics literature 
\cite{Lazaroiu:2001nm,AK}, 
we will not provide a comprehensive review here. Rather we will 
recall some basic elements needed for our construction. 

In order to extend this computational framework to orientifold models, 
we have to find an off-shell cochain model equipped with an orientifold 
projection and a compatible differential algebraic structure. 
We made a first 
step in this direction in the previous section by giving a categorical 
formulation of the orientifold projection. In section \ref{s:3.1} we will 
refine this construction, obtaining the desired cochain model. 

Having constructed a suitable cochain model, the computation of 
the superpotential follows the same pattern as in the absence of 
the orientifold projection. A notable distinction resides in the  
occurrence of $L_\infty$ instead of $A_\infty$  
structures, since the latter are not compatible with the involution. 
The final result obtained in section \ref{s:3.2} is that the orientifold 
superpotential can be obtained by evaluating the superpotential 
of the underlying unprojected theory on invariant field configurations.

\subsection{Cochain Model and Orientifold Projection}\label{s:3.1}

Suppose $\me$ is a locally free complex on $X$, and that it is left invariant by the 
parity functor. This means that $\me$ and 
$P(\me)$ are isomorphic in the derived category, and we choose such an isomorphism 
\be\label{eq:quasisom}
 \psi \colon \me \to P(\me).
\ee
Although in general $\psi$ is not a map of complexes, it can be chosen so 
in most practical situations, including all cases studied in this paper. 
Therefore we will assume from now on that $\psi$ is a quasi-isomorphism 
of complexes:
\be\label{eq:psiform} 
\xymatrix{
& \cdots \ar[r]&  E_{m-i+1} \ar[rr]^{d_{m-i+1}}\ar[d]^{\psi_{m-i+1}} && 
E_{m-i}\ar[rr]^{d_{m-i}}\ar[d]^{\psi_{m-i}} &&  E_{m-i-1}\ar[d]^{\psi_{m-i-1}} 
\ar[r]& \cdots\\
& \cdots \ar[r]&  P(E_{i-1}) \ar[rr]^{P(d_{i})} & &
P(E_{i})\ar[rr]^{ P(d_{i+1})} & & 
P(E_{i+1}) \ar[r]
& \cdots\\}
\ee
We have written \eqref{eq:psiform} 
so that the terms in the same column have the same degree since 
$\psi$ is a degree zero morphism. The degrees of the three columns from left to right 
are $i-m-1,\ i-m$ and $i-m+1$. 
For future reference, note that the 
quasi-isomorphism $\psi$ induces a quasi-isomorphism of cochain complexes 
\be\label{eq:cochaineq}
\psi_* \colon \CC(P(\me),\me) \to \CC(P(\me),P(\me)),\qquad
f \mapsto \psi\comp f. 
\ee

The problem we are facing in the construction of a viable cochain 
model resides in the absence of a natural orientifold projection 
on the cochain space $\CC(\me,\me)$. 
$P$ maps $\CC(\me,\me)$ to 
$\CC(P(\me),P(\me))$, which is not identical to $\CC(\me,\me)$. 
How can we find a natural orientifold projection on a given
off-shell cochain model? 

Since $\me$ and $P(\me)$ are quasi-isomorphic, one can equally 
well adopt the morphism space 
\[
\CC(P(\me),\me) = {\mathrm{Mor}}_{\textrm{Pre-Tr}(\wCC)}(P(\me),\me)
\]
as an off-shell cochain model. As opposed to $\CC(\me,\me)$, 
this morphism space has a natural induced involution
defined by the composition
\be\label{eq:cochaininv}
\xymatrix@=13mm@1{
& \CC(P(\me), \me) \ar[r]^-P & *+{\,\CC(P(\me),P^2(\me))}
\ar[r]^-{J_\ast}& *+++{\CC(P(\me),\me)}.}
\ee
where $J$ is  the isomorphism in \eqref{eq:paractK}.
Therefore we will do our superpotential computation 
in the cochain model $\CC(P(\me),\me)$, as opposed to $\CC(\me,\me)$, 
which is used in \cite{AK}. 

This seems to lead us to another puzzle, since a priori there is 
no natural associative algebra structure on $\CC(P(\me),\me)$. 
One can however define one using the quasi-isomorphism 
\eqref{eq:quasisom}. Given 
\[ 
f^p_{q,k} \in A^{0,p}(\sHom_X(P(E_k),E_{m-k-q}))\qquad 
g^r_{s,l} \in A^{0,r}(\sHom_X(P(E_l),E_{m-l-s}))
\]
we define 
\be\label{eq:psiprod}
g^{r}_{s,l}\star_\psi f^p_{q,k} =\left\{ 
\begin{array}{ll}
(-1)^{sp} g^{r}_{s,l}\cdot \psi_{m-k-q} \cdot f^p_{q,k} & \mbox{for $l=k+q$}    \\
0   & \mbox{otherwise}.
\end{array}     \right.
\ee
where $\cdot$ denotes exterior multiplication of bundle valued
differential forms. 

With this definition, the map \eqref{eq:cochaineq} becomes 
a quasi-isomorphism of DGAs. 
The sign $(-1)^{sp}$ in \eqref{eq:psiprod} 
is determined by the sign rule \eqref{eq:signruleA} for composition 
of morphisms in $\wCC$. 
This construction 
has the virtue that it makes both the algebra 
structure and the orientifold projection manifest. Note that 
the differential $Q$ satisfies the graded Leibniz rule with respect 
to the product $\star_\psi$ because $\psi$ is a $Q$-closed 
element of $\CC(P(\me),\me)$ of degree zero. 

Next we check two compatibility conditions between 
the involution \eqref{eq:cochaininv} and the DGA structure. 

\begin{lemma}
For any cochain $f \in \CC(P(\me),\me)$
\be\label{eq:consistA} 
J_*P(Q(f)) = Q(J_*P(f)).
\ee
\end{lemma}
\begin{proof}
Using equation \eqref{eq:paractI}, the explicit expression for the 
differential $Q$ acting on a 
homogeneous element $f^p_{q,k}$ as above is 
\[ 
Q(f^p_{q,k}) = \dbar_{P(E_k) E_{m-k-q}} (f^p_{q,k}) 
+ (d_{m-k-q} \comp f^{p}_{q,k})_\wCC -(-1)^{p+q} 
(f^{p}_{q,k}\comp P(d_k))_\wCC.
\]
According to equation \eqref{eq:paractH}, we have 
\be\label{eq:consistB}
\begin{aligned} 
P(Q(f^p_{q,k})) = & \dbar_{P(E_{m-k-q})P^2(E_k)}(P(f^p_{q,k})) + 
(P^2(d_k)\comp P(f^p_{q,k}))_\wCC\cr
&  -(-1)^{|P(f)|} 
(P(f^p_{q,k})\comp P(d_{m-k-q}))_\wCC \cr 
\end{aligned}
\ee
The commutative diagram \eqref{eq:paractK} shows that 
\[
J\comp P^2(d_k) = d_{k} \comp J. 
\] 
Then, equation \eqref{eq:consistB} yields 
\[ 
\begin{aligned}
J_*P(Q(f^p_{q,k})) = & \dbar_{P(E_{m-k-q})E_k}(J_*P(f^p_{q,k}))
+ (d_k\comp J_*P(f^p_{q,k}))_\wCC \cr
& -(-1)^{|f|} 
(J_*P(f^p_{q,k}) \comp P(d_{m-k-q}))_\wCC
\end{aligned} 
\]
which proves \eqref{eq:consistA}. 
\end{proof}

\begin{lemma}
For any two elements $f,g\in\CC(P(\me),\me)$
\be\label{eq:consistC}
J_*P(g\star_\psi f) = -(-1)^{|f||g|} J_*P(f) \star_\psi J_*P(g).
\ee
\end{lemma}
\begin{proof}
Written in terms of homogeneous elements, \eqref{eq:consistC} 
reads 
\be\label{eq:consistCA} 
J_*P(g^{r}_{s,l}\star_{\psi} f^p_{q,k}) = -(-1)^{(r+s)(p+q)}
J_*P(f^p_{q,k}) \star_{\psi} J_*P(g^r_{s,l})
\ee
where $l=k+q$. 
Using equations \eqref{eq:paractD}, \eqref{eq:psiprod} and the definition of 
\eqref{eq:paractK} of $J$, we compute 
\[ 
\begin{aligned} 
J_*P(g^{r}_{s,l}\star_\psi f^p_{q,k}) 
& = (-1)^{(m-s-l)(m+1)+\omega} (-1)^{(s+q)(m-s-l)+1} (-1)^{sp}\,
\sigma^*(g^r_{s,l}\cdot \psi_{m-k-q}\cdot f^p_{q,k})^\roof \\
& = (-1)^{(m-s-l)(m+1)+\omega} 
(-1)^{(s+q)(m-s-l)+1} (-1)^{sp} (-1)^{rp}\\ 
& \ \quad \sigma^*(f^p_{q,k})^\roof \cdot
\sigma^*(\psi_{m-k-q}^\roof)\cdot 
\sigma^*(g^r_{s,l})^\roof \\
& = (-1)^{(m-s-l)(m+1)+\omega} (-1)^{(s+q)(m-s-l)+1} (-1)^{sp} 
(-1)^{rp}\\
& \quad\ (-1)^{(m-k-q)(m+1)+\omega} 
(-1)^{q(q+k-m)+1}(-1)^{(m-s-l)(m+1)+\omega} 
(-1)^{s(s+l-m)+1}\\
&\quad\  J_{\ast}P(f^p_{q,k})\cdot
\sigma^*(\psi_{m-k-q}^\roof)
\cdot J_{\ast}P (g^r_{s,l})
\end{aligned}
\]
\[
\begin{aligned}
-(-1)^{(r+s)(p+q)}J_*P(f^p_{q,k}) \star_{\psi} J_*P(g^r_{s,l})
&= -(-1)^{(r+s)(p+q)}(-1)^{qr}J_*P(f^p_{q,k})\cdot\psi_{l}\cdot J_*P(g^r_{s,l})
\end{aligned}
\]
These expressions are in agreement with equation \eqref{eq:consistCA}
if and only if $\psi$ satisfies a symmetry condition of the form 
\be\label{eq:psicond}
J^{\ast}P(\psi_{m-l}) = - \psi_{l} \quad\Leftrightarrow\quad
\sigma^{\ast}(\psi_{m-l})^{\roof} = (-1)^{(m+1)l+\omega} \psi_{l}
\ee
\end{proof}

We saw in the last proof that compatibility of the orientifold projection
with  the algebraic structure imposes the condition 
\eqref{eq:psicond} on  $\psi$. From now on we  
assume this condition to be satisfied. Although we do not know a general
existence result for a quasi-isomorphism satisfying \eqref{eq:psicond}, 
we will show that such a choice is possible in all the examples considered 
in this paper. We will also see that symmetry 
of $\psi$, which is determined by $\omega=0,1$ 
in \eqref{eq:psicond}, determines
whether the orientifold projection is of type $SO$ or $Sp$. 

Granting such a quasi-isomorphism, it follows that the cochain space
$\CC(P(\me),\me)$ satisfies all the conditions required for   
the computation of the superpotential, which is the subject of the next 
subsection. 

\subsection{The Superpotential}\label{s:3.2}

In the absence of an orientifold projection, the computation of the 
superpotential can be summarized as follows \cite{Fukaya}. 
Suppose we are searching for formal deformations of the differential 
$Q$ of the form 
\be\label{eq:formdefA}
Q_{\mathrm{def}}=Q + f_1(\phi)+f_2(\phi)+f_3(\phi)+\ldots 
\ee
where $$f_1(\phi)=\phi$$ is a cochain of degree one, which represents 
an infinitesimal deformation of $Q$. The terms $f_k(\phi)$, for
$k\geq 2$, are homogeneous polynomials of degree $k$ in $\phi$ 
corresponding to higher order deformations. 
We want to impose the integrability condition 
\be\label{eq:intcond} 
(Q_{\mathrm{def}})^2 =0 
\ee
order by order in $\phi$. In doing so one encounters certain 
obstructions, which are 
systematically encoded in a minimal $A_\infty$ model of the 
DGA $\CC(P(\me),\me)$. 
The superpotential is essentially a primitive 
function for the obstructions, and exists under certain 
cyclicity conditions. 

In the orientifold model we have to solve a similar deformation 
problem, except that now the deformations
of $Q$ have to be invariant under the orientifold action. We will explain below that this 
is equivalent to the construction of a minimal $L_\infty$ model. 

Let us first consider the integrability conditions \eqref{eq:intcond}
in more detail in the absence of  orientifolding.
Suppose we are given an associative $\IZ$-graded  DGA  
$(\CC,Q,\cdot)$, and let $H$ denote the cohomology of $Q$. 
In order to construct an $A_\infty$ structure on $H$ we need the following data 

$(i)$ A $\IZ$-graded linear subspace $\CH\subset \CC$ isomorphic to the 
cohomology of $Q$. In other words, $\CH$ is spanned in each degree by
representatives of the cohomology classes of $Q$. 

$(ii)$ A linear map $\eta: \CC\to \CC[-1]$ mapping $\CH$ to itself such that 
\be\label{eq:proj}
\Pi = {\mathbb I} - [Q,\eta] 
\ee
is a projector $\Pi\colon \CC\to \CH$, where $[\quad , \quad]$ is the 
graded commutator.  
Moreover, we assume that the following conditions are satisfied 
\be\label{eq:mincond}
\eta|_{\CH}=0\qquad \eta^2=0. 
\ee
\noindent 
Using  the data $(i),\ (ii)$ one can develop a recursive approach to obstructions 
in the deformation theory of $Q$ \cite{Fukaya}. The integrability condition 
\eqref{eq:intcond} yields 
\be\label{eq:intcondB} 
\begin{aligned} 
\sum_{n=1}^\infty \left[Q(f_n(\phi))+B_{n-1}(\phi)\right]=0
\end{aligned}
\ee 
where 
\[ 
\begin{aligned}
B_0&=0\cr
B_{n-1}& = \phi f_{n-1}(\phi) + f_{n-1}(\phi) \phi + 
\mathop{\sum_{k+l=n}}_{k,l\geq 2} f_k(\phi) f_l(\phi),\quad n\geq2 
\cr 
\end{aligned}
\]
Using equation \eqref{eq:proj}, we can rewrite equation \eqref{eq:intcondB}
as 
\be\label{eq:intcondC} 
\begin{aligned} 
& \sum_{n=1}^\infty 
\left[Q(f_n(\phi))+([Q,\eta]+\Pi)B_{n-1}(\phi)\right]=0.\cr
\end{aligned}
\ee
We claim that the integrability condition \eqref{eq:intcondB} can be solved recursively 
\cite{Fukaya} provided that 
\be\label{eq:intcondD} 
\sum_{n=1}^\infty \Pi(B_{n-1}) =0.
\ee
To prove this claim, note that if the condition \eqref{eq:intcondD} 
is satisfied, equation \eqref{eq:intcondC} becomes 
\be\label{eq:intcondE} 
\begin{aligned} 
& \sum_{n=1}^\infty 
\left(Q(f_n(\phi))+[Q,\eta]B_{n-1}(\phi)\right)=0.\cr
\end{aligned}
\ee
This equation can be solved by setting recursively  
\be\label{eq:intcondF}
f_n(\phi) = - \eta(B_{n-1}(\phi)).
\ee
One can show that this is a solution to \eqref{eq:intcondF} 
by proving inductively that 
\[
Q(B_{n}(\phi))=0.
\]
In conclusion, the obstructions to the integrability condition
\eqref{eq:intcondB} are encoded in the formal series 
\be\label{eq:obstrA}
\sum_{n=2}^\infty \Pi\bigg(\phi f_{n-1}(\phi) + f_{n-1}(\phi) \phi + 
\mathop{\sum_{k+l=n}}_{k,l\geq 2} f_k(\phi) f_l(\phi)\bigg)  
\ee
where the $f_n(\phi)$, $n\geq 1$, are determined recursively 
by \eqref{eq:intcondF}. 

The algebraic structure emerging from this construction is 
a minimal $A_\infty$ structure for the DGA $(\CC,Q)$ 
\cite{Kad,Merkulov}. 
\cite{Merkulov} constructs an 
$A_\infty$ structure by defining the linear maps
\[ 
\lambda_n \colon \CC^{\otimes n} \to \CC[2-n], \qquad n\geq 2 
\] 
recursively
\be\label{eq:minmodA} 
\begin{aligned} 
\lambda_n(c_1,\ldots, c_n) = & 
(-1)^{n-1}(\eta\lambda_{n-1}(c_1,\ldots,c_{n-1}))\cdot c_n
-(-1)^{n|c_1|} c_1\cdot \eta\lambda_{n-1}(c_2,\ldots, c_n) \cr
& -\mathop{\sum_{k+l=n}}_{k,l\geq 2} 
(-1)^{r} [\eta\lambda_k(c_1,\ldots,c_k)]\cdot 
[\eta\lambda_l(c_{k+1},\ldots,c_n)]\cr
\end{aligned}
\ee
where $|c|$ denotes the degree of an element $c\in \CC$, and 
\[
r = k+1 +(l-1)(|c_1|+\ldots+|c_k|).
\] 
Now define the linear maps 
\[ 
{\mathfrak m}_n \colon \CH^{\otimes n} \to \CH[2-n], \qquad n \geq 1 
\] 
by 
\be\label{eq:minmodB} 
\bal 
{\mathfrak m}_1 & = \eta \cr
{\mathfrak m}_n & = \Pi \lambda_n. 
\eal
\ee
The products \eqref{eq:minmodB} 
define an $A_\infty$ structure on $\CH\simeq H$. If the conditions 
\eqref{eq:mincond} are satisfied, this $A_\infty$ structure is a 
minimal model for the DGA $(\CC,Q,\cdot)$. The products ${\mathfrak m}_n$, $n\geq 2$
agree up to sign with the obstructions $\Pi(B_n)$ found above. 

The products ${\mathfrak m}_n$  determine the local equations 
of the D-brane moduli space, which in physics language are called 
F-term equations. If 
\[
\phi = \sum_{i=1}^{\rm{dim}(\CH)} \phi^i u_i 
\] 
is an arbitrary cohomology element written in terms of some 
generators $\{u_i\}$, the F-term equations are 
\be\label{eq:FtermA}
\sum_{n=2}^\infty (-1)^{n(n+1)/2}{\mathfrak m}_n(\phi^{\otimes n}) =0.
\ee
If the products are cyclic, these equations admit a primitive 
\be\label{eq:suppotA}
W = \sum_{n=2}^\infty \frac{(-1)^{n(n+1)/2}}{n+1}
\langle \phi, {\mathfrak m}_n(\phi^{\otimes n})\rangle 
\ee
where 
\[
\langle\quad , \quad \rangle \colon \CC \to \IC
\] 
is a bilinear form on $\CC$ compatible with the DGA structure. 
The cyclicity property reads
\[
\langle c_1,{\mathfrak m}_n(c_2,\ldots,c_{n+1})\rangle = 
(-1)^{n|c_2|+1}\langle c_2, {\mathfrak m}_n(c_3,\ldots,c_{n+1},c_1)\rangle.
\]

Let us now examine the above deformation problem in the presence of 
an orientifold projection. Suppose we have an involution 
$\tau \colon \CC\to \CC$ such that the following conditions are satisfied
\be\label{eq:invcondA} 
\begin{aligned} 
\tau(Q(f)) & = Q(\tau(f)) \cr
\tau(fg) & = -(-1)^{|f||g|}\tau(g)\tau(f) \cr
\end{aligned} 
\ee 
As explained below equation \eqref{eq:intcond}, 
in this case we would like to study deformations 
\[
Q_{def} = Q + f_1(\phi) + f_2(\phi) + \ldots  
\]
of $Q$ such that 
\be\label{eq:invcondB}
\tau(f_n(\phi)) = f_n(\phi) 
\ee
for all $n\geq 1$.

In order to set this problem in the proper algebraic context, note that 
the DG algebra $\CC$ decomposes into a direct sum of $\tau$-invariant 
and anti-invariant parts
\be\label{eq:decomp}
\CC \simeq \CC^+ \oplus \CC^-.
\ee
There is a similar decomposition 
\be\label{eq:decompB} 
H = H^+ \oplus H^- 
\ee 
for the $Q$-cohomology. 

Conditions \eqref{eq:invcondA} imply that $Q$ preserves $\CC^\pm$, but the 
associative algebra product is not compatible with the decomposition 
\eqref{eq:decomp}. There is however another algebraic structure which 
is preserved by $\tau$, namely the graded commutator 
\be\label{eq:grdcomm} 
[f,g] = fg - (-1)^{|f||g|} \, gf.
\ee
This follows immediately from the second equation in \eqref{eq:invcondA}. 
The graded commutator \eqref{eq:grdcomm} defines a differential graded 
Lie algebra structure on $\CC$. By restriction, it also defines a 
DG Lie algebra structure on the invariant part $\CC^+$. 
In this context our problem reduces to the deformation theory of 
the restriction $Q^+ = Q|_{\CC^+}$ as a differential operator 
on $\CC^+$.  

Fortunately, this problem can be treated by analogy with the previous 
case, except that we have to replace $A_\infty$ structures by $L_\infty$ 
structures, see for example \cite{MerkulovL,Lazaroiu:2001nm,Fukaya}.
In particular, the obstructions to the deformations of $Q^+$ can be 
systematically encoded in a minimal $L_\infty$ model, and one can 
similarly define a superpotential if certain cyclicity conditions 
are satisfied. 

Note that any associative DG algebra can be naturally endowed with a 
DG Lie algebra structure using the graded commutator \eqref{eq:grdcomm}. 
In this case, the $A_\infty$ and the $L_\infty$ approach to the
deformation of $Q$ are equivalent \cite{Lazaroiu:2001nm} 
and they yield the same superpotential. However, the $L_\infty$ approach 
is compatible with the involution, while the $A_\infty$ approach is not.

To summarize this discussion, we have a DG Lie algebra on $\CC$ which 
induces a DG Lie algebra of $Q$. The construction of a minimal 
$L_\infty$ model for $\CC$ requires the same data $(i),\ (ii)$ as 
in the case of a minimal $A_\infty$ model, and yields the same 
F-term equations, and the same superpotential. In order to determine 
the F-term equations and superpotential for the invariant part 
$\CC^+$ we need again a set of data $(i),\ (ii)$ as described 
above \eqref{eq:proj}. This data can be naturally obtained by 
restriction from $\CC$ provided that the propagator $\eta$
in equation \eqref{eq:proj} 
can be chosen compatible with the involution $\tau$ i.e 
\[
\tau(\eta(f)) = \eta(\tau(f)). 
\]
This condition is easily satisfied in geometric situations, 
hence we will assume that this is the case from now on. 
Then the propagator $\eta^+: \CC^+ \to \CC^+[-1]$ 
is obtained by restricting $\eta$ to the invariant part 
$\eta^+ = \eta|_{\CC^+}$. 
Given this data, we construct a minimal $L_\infty$ model for the 
DGL algebra $\CC^+$, which yields F-term equations and, if the 
cyclicity condition is satisfied, a superpotential $W^+$. 

\begin{thm}\label{c:20}
The superpotential $W^+$ obtained by constructing the 
minimal $L_\infty$ model for the DGL $\CC^+$ is equal to the restriction 
of the superpotential $W$ corresponding to $\CC$ evaluated on 
$\tau$-invariant field configurations:
\be\label{eq:invsup}
W^+ = W|_{H^+}. 
\ee
\end{thm}
In the remaining part of this section we will give a formal argument for 
this claim. 
According to \cite{MerkulovL}, the data $(i), \ (ii)$ above equation 
\eqref{eq:proj}
also determines an $L_\infty$ structure on $\CH$ as follows. 
First we construct a series of linear maps 
\[
\rho_n \colon \CC^{\otimes n} \to \CC[2-n], \qquad n\geq 2
\] 
by anti-symmetrizing (in the graded sense) the maps \eqref{eq:minmodA}. 
That is the recursion relation becomes 
\be\label{eq:minmodC}
\begin{aligned}
\rho_n(c_1,\ldots,c_n) = & \sum_{\sigma\in Sh(n-1,1)}
(-1)^{n-1+|\sigma|}e(\sigma)[\eta\rho_{n-1}(c_{\sigma(1)},\ldots,c_{\sigma(n-1)}), 
c_{\sigma(n)}] \cr 
& -\sum_{\sigma\in Sh(1,n)}(-1)^{n|c_1|+|\sigma|}e(\sigma)
[c_1,\eta\rho_{n-1}(c_{\sigma(2)}]\cr
& -\mathop{\sum_{k+l=n}}_{k,l\geq 2} \sum_{\sigma\in Sh(k,n)}(-1)^{r+|\sigma|}e(\sigma) 
[\eta\rho_k(c_{\sigma(1)},\ldots,c_{\sigma(k)}), 
\eta\rho_l(c_{\sigma(k+1)},\ldots,c_{\sigma(n)})]\cr
\end{aligned}
\ee
where $Sh(k,n)$ is the set of all
permutations $\sigma\in S_n$ such that 
\[ 
\sigma(1)<\ldots < \sigma(k) \quad \mathrm{and} \quad 
\sigma(k+1) < \ldots < \sigma(n)
\]
and $|\sigma|$ is the signature of a permutation $\sigma\in S_n$.
The symbol $e(\sigma)$ denotes the Koszul sign defined by 
\[
c_{\sigma(1)}\wedge \ldots \wedge c_{\sigma(n)} = 
(-1)^{|\sigma|} e(\sigma) c_1 \wedge \ldots \wedge c_n.
\]
Then we define the $L_\infty$ products 
\[
{\mathfrak l}_n \colon \CH^{\otimes n} \to \CH 
\]  
by 
\be\label{eq:minmodD}
{\mathfrak l}_1   = \eta ,\qquad
{\mathfrak l}_n  = \Pi \rho_n .
\ee
One can show that these products satisfy a series of higher 
Jacobi identities analogous to the defining associativity 
conditions of $A_\infty$ structures. If the conditions \eqref{eq:mincond} 
are also satisfied, the resulting $L_\infty$ structure is a minimal 
model for the DGL algebra $\CC$. 

Finally, note that the $A_\infty$ products \eqref{eq:minmodB} and 
the $L_\infty$ products \eqref{eq:minmodD} are related by 
\be\label{minmodrel} 
{\mathfrak l}_n(c_1,\ldots,c_n) = \sum_{\sigma\in S_n} (-1)^{|\sigma|} e(\sigma) 
{\mathfrak m}_n(c_{\sigma(1)}, \ldots, c_{\sigma(n)}). 
\ee
In particular, one can rewrite the F-term equations \eqref{eq:FtermA} 
and the superpotential \eqref{eq:suppotA} in terms of $L_\infty$ 
products \cite{Lazaroiu:2001nm,Fukaya}. 

The construction of the minimal $L_\infty$ model of the invariant part 
$\CC^+$ is analogous. Since we are working under assumption that the 
propagator $\eta^+$ is the restriction of $\eta$ to $\CC^+$, 
it is clear that the linear maps 
$\rho_n^+(c_1,\ldots,c_n)$
are also equal to the restriction $\rho_n|_{(\CC^+)^n}$. 
The same will be true for the products ${\mathfrak l}_n^+$, 
i.e. 
\[ 
{\mathfrak l}_n^+ = {{\mathfrak l}_n}|_{(H^+)^n}.
\] 
Therefore the F-term equations and the superpotential in the orientifold
model can be obtained indeed by restriction to the invariant part. 

Now that we have the general machinery at hand, we can turn to
concrete examples of superpotential computations. 
  
\section{Computations for Obstructed Curves}\label{s:4}

In this section we perform detailed  computations of the superpotential
for D-branes wrapping holomorphic curves in Calabi-Yau orientifolds. 

So far we have relied on the Dolbeault cochain 
model, which serves as a good conceptual framework 
for our constructions. However, a $\check{\mathrm{C}}$ech cochain 
model is clearly preferred for computational purposes 
\cite{AK}. The simple prescription found above for the 
orientifold superpotential allows us to switch 
from the Dolbeault to the $\check{\mathrm{C}}$ech
model with little effort. Using the same definition 
for the action of the orientifold projection $P$ on 
locally free complexes $\me$, we will adopt a
cochain model of the form 
\be\label{eq:cechA}
\CC(P(\me),\me) = \check{C}(\mfu, {\sHom}_X(P(\me),\me))
\ee
where $\mfu$ is a fine open cover of $X$. 
The differential $Q$ is given by 
\be\label{eq:cechB}
Q(f) = \delta(f) +(-1)^{c(f)} \mathfrak{d}(f)
\ee
where $\delta$ is the $\check{\mathrm{C}}$ech differential, 
$\mathfrak{d}$ is the
differential of the local Hom complex and $c(f)$ is the 
$\check{\mathrm{C}}$ech degree of $f$.

In order to obtain a well-defined involution on the complex 
\eqref{eq:cechA}, 
we have to choose the open cover $\mfu$ so that the holomorphic involution 
$\sigma\colon X\to X$ maps any open set $U\in \mfu$ isomorphically to another
 open set $U_{s(\alpha)}\in \mfu$, where $s$ is an involution on the set of 
indices $\{\alpha\}$. Moreover, the holomorphic involution 
should also be compatible with intersections. That is, if $U_\alpha,
U_\beta\in \mfu $ 
are mapped to $U_{s(\alpha)}, U_{s(\beta)}\in \mfu$ then 
$U_{\alpha\beta}$ should be mapped isomorphically to 
$U_{s(\alpha)s(\beta)}$. Analogous properties should 
hold for arbitrary multiple intersections. Granting such a choice of a fine open cover, we have a natural involution 
$J_{\ast}P$ acting on the cochain complex \eqref{eq:cechA}, defined as in  \eqref{eq:cochaininv}.

According to the prescription derived in the previous section,
the orientifold superpotential can be obtained by applying 
the computational scheme of \cite{AK} to invariant $Q$-cohomology 
representatives. 
Since the computation depends only on the infinitesimal 
neighborhood of the curve, it suffices to consider 
local Calabi-Yau models as in \cite{AK}. We will consider two representative cases, 
namely obstructed $(0,-2)$ curves and local conifolds, i.e., $(-1,-1)$ curves. 

\subsection{Obstructed $(0,-2)$ Curves in $O5$ Models} 

In this case, the local Calabi-Yau $X$ 
can be covered by two coordinate patches $(x,y_1,y_2)$, $(w,z_1,z_2)$ 
with transition functions 
\be\label{eq:exoneA}
\begin{aligned} 
w&=x^{-1} \cr
z_1& =x^2y_1+ xy_2^n\cr 
z_2&=y_2. \cr
\end{aligned}
\ee
The $(0,-2)$ curve is given by the equations 
\be\label{eq:exoneB}
C\colon \quad  y_1=y_2=0\qquad \mbox{resp.} \quad z_1=z_2=0
\ee
in the two patches. 
The holomorphic involution acts as
\be\label{eq:exoneC} 
\begin{aligned} 
(x,y_1,y_2) & \mapsto (x, -y_1,-y_2)\cr
(w,z_1,z_2) & \mapsto (w, -z_1,-z_2) \cr
\end{aligned} 
\ee 
This is compatible with the transition functions if and only if 
$n$ is odd. We will assume that this is the case from now on. 
Using \eqref{eq:invconfB}, the Chan-Paton bundles
\be\label{eq:exoneD} 
V_{N} = \co_C(-1)^{\oplus N}.
\ee
define invariant D-brane configurations under the orientifold projection.

The on-shell open string states are in one-to-one correspondence with 
elements of the global Ext group $\Ext^1(i_* V_N,i_* V_N)$. 
Given two bundles $V,W$ supported on 
a curve $i\colon C\hookrightarrow X$, there is a spectral sequence \cite{KS}
\be\label{eq:specseq}
E_2^{p,q}=H^p(C,V^\roof\otimes W\otimes \Lambda^q N_{C/X}) \ \Rightarrow \ 
\mathrm{Ext}^{p+q}_X(i_* V,i_* W)
\ee
which degenerates at $E_2$. This yields 
\[
\Ext^1(i_*\co_C(-1),i_*\co_C(-1)) \simeq H^0(C,N_{C/X})= \IC ,
\]
since $N_{C/X} \simeq \co_C \oplus \co_C(-2)$. 
Therefore a D5-brane with multiplicity $N=1$ has a single normal 
deformation. For higher multiplicity, the normal deformations 
will be parameterized by an $(N\times N)$ complex matrix.

In order to apply the computational algorithm developed 
in section \ref{s:3} we have to find a locally free resolution $\me$ 
of $i_*\co_C(-1)$ and an explicit generator of  
$$\Ext^1(i_*\co_C(-1),i_*\co_C(-1))\simeq \Ext^1(P(\me),\me)$$
in the cochain space $\check{C}(\mfu, \sHom(P(\me),\me))$. 
We  take $\me$ to be the locally free resolution from 
\cite{AK} multiplied by $\co_C(-1)$, i.e.,
\begin{equation}\label{eq:resolF}
\xymatrix@C=8mm@M=1mm{
0 \ar[r] &
\co(-1) \ar[rr]^{\left(\begin{smallmatrix} y_{2} \\ -1 \\ x \end{smallmatrix}\right)} & &
{\begin{matrix} \co(-1) \\ \oplus \\ \co \\ \oplus \\ \co \end{matrix}} 
\ar[rrrr]^{\left(\begin{smallmatrix} 1 & y_{2} & 0 \\ -x & 0 & y_{2} \\
-y_{2}^{n-1} & -s & -y_{1} \end{smallmatrix}\right)} & & & &
{\begin{matrix} \co \\ \oplus \\ \co \\ \oplus \\ \co(-1) 
\end{matrix}} \ar[rrr]^{\left(\begin{smallmatrix} s & y_{1} & y_{2} 
\end{smallmatrix}\right)} & & &
\co(-1)\\ 
}
\end{equation}
The quasi-isomorphism $\psi\colon \me\to P(\me)$ is given by 
\begin{equation}\label{eq:quasisomB}
\xymatrix@C=8mm@M=1mm{
{\begin{matrix} \co(-1) \\ \oplus \\ \co^{\oplus 2} \end{matrix}} 
\ar[rrrr]^{\left(\begin{smallmatrix} 1 & y_{2} & 0 \\ -x & 0 & y_{2} \\ 
-y_{2}^{n-1} & -s & -y_{1} \end{smallmatrix}\right)} 
\ar[dd]_{\left(\begin{smallmatrix} 0 & x & 1 \end{smallmatrix}\right)} 
& & & &
{\begin{matrix} \co^{\oplus 2} \\ \oplus \\ \co(-1) \end{matrix}} 
\ar[rrrr]^{\left(\begin{smallmatrix} s & y_{1} & y_{2}
    \end{smallmatrix}\right)} 
\ar[dd]_{\left(\begin{smallmatrix} 0 & y_{2}^{n-1} & -x \\ -y_{2}^{n-1} & 0 &
-1 \\ x & 1 & 0 \end{smallmatrix}\right)} & & & &
\co(-1) \ar[dd]^{\left(\begin{smallmatrix} 0 \\ x \\ 1
    \end{smallmatrix}\right)} 
\\ \\
\co(1) \ar[rrrr]^{\left(\begin{smallmatrix} s \\ y_{1} \\ y_{2} 
\end{smallmatrix}\right)} & & & &
{\begin{matrix} \co^{\oplus 2} \\ \oplus \\ \co(1) \end{matrix}} 
\ar[rrrr]^{\left(\begin{smallmatrix} 1 & -x & -y_{2}^{n-1} \\ -y_{2} & 0 & s 
\\ 0 & -y_{2} & y_{1} \end{smallmatrix}\right)} & & & &
{\begin{matrix} \co(1) \\ \oplus \\ \co^{\oplus 2} \end{matrix}} 
}
\end{equation}
Note that $\psi$ satisfies the symmetry condition \eqref{eq:psicond} with
$\omega = 0$, which in this case reduces to
\be\label{eq:quasisomBcond}
\sigma^*(\psi_{2-l})^\roof =(-1)^{l}\psi_l.
\ee

We are searching for a generator ${\mathsf c}\in
\check{C}(\mfu,\sHom(P(\me),\me))$ 
of the form 
$\msc = \msc^{1,0} + \msc^{0,1} $
for two homogenous elements 
\[
\msc^{p,1-p} \in \check{C}^p(\mfu,\sHom^{1-p}(P(\me),\me)),\qquad 
p=0,1.
\]
The cocycle condition $Q(\msc)=0$ is equivalent to 
\begin{equation}\label{eq:closureA}
\begin{aligned}
\mathfrak{d}\mathsf{c}^{0,1} & = \delta\mathsf{c}^{1,0} = 0\cr
Q(\mathsf{c}^{0,1} + \mathsf{c}^{1,0}) & 
= \delta\mathsf{c}^{0,1} - \mathfrak{d}\mathsf{c}^{1,0} = 0\cr
\end{aligned}
\end{equation}
A solution to these equations is given by 
\begin{equation}\label{eq:genA}
\begin{xy} <1em,0em>:
(0,0)*\xybox{\xymatrix@C=8mm@M=1mm{
\co(1) \ar[rr] 
\ar[dd]^{\left(\begin{smallmatrix} x^{-1} \\ 0 \\ 0 \end{smallmatrix}\right)_{01}} & &
{\begin{matrix} \co^{\oplus 2} \\ \oplus \\ \co(1) \end{matrix}} 
\ar[rr] \ar[dd]^{\left(\begin{smallmatrix} 0 & 0 & 0 \\ 0 & 0 & 0 \\ 0 & 0 
& -x^{-1}y_{2}^{n-2} \end{smallmatrix}\right)_{01}} & &
{\begin{matrix} \co(1) \\ \oplus \\ \co^{\oplus 2} \end{matrix}} 
\ar[dd]^{\left(\begin{smallmatrix} x^{-1} & 0 & 0 \end{smallmatrix}\right)_{01}} \\ \\
{\begin{matrix} \co(-1) \\ \oplus \\ \co^{\oplus 2} \end{matrix}} \ar[rr] & &
{\begin{matrix} \co^{\oplus 2} \\ \oplus \\ \co(-1) \end{matrix}} \ar[rr] & &
\co(-1)
}},
(-4,-5.5)*{\mathsf{c}^{1,0} := }
\end{xy}
\end{equation}
\[
\begin{xy} <1em,0em>:
(0,0)*\xybox{\xymatrix@C=8mm@M=1mm{
\co(1) \ar[rr] \ar[dd]^{\left(\begin{smallmatrix} 0 \\ -1 \\ 0 
\end{smallmatrix}\right)_{0} + \left(\begin{smallmatrix} 1 \\ 0 \\ 0 
\end{smallmatrix}\right)_{1}} & &
{\begin{matrix} \co^{\oplus 2} \\ \oplus \\ \co(1) \end{matrix}} 
\ar[dd]^{\left(\begin{smallmatrix} 0 & 1 & 0 \end{smallmatrix}\right)_{0} + 
\left(\begin{smallmatrix} -1 & 0 & 0 \end{smallmatrix}\right)_{1}} \\ \\
{\begin{matrix} \co^{\oplus 2} \\ \oplus \\ \co(-1) \end{matrix}} \ar[rr] & &
\co(-1)
}},
(-4,-5.5)*{\mathsf{c}^{0,1} := }
\end{xy}
\]
These satisfy the symmetry conditions 
\be\label{eq:genB} 
J_{\ast}P(\msc^{p,1-p}) = -(-1)^\omega \msc^{p,1-p},\qquad p=0,1,
\ee

For multiplicity $N > 1$, we have the locally free resolution 
$\me_N = \me \otimes \IC^N$.
The quasi-isomorphism $\psi_N \colon \me_N \to P(\me_N)$ is of the 
form $\psi_N = \psi \otimes M $,
where $M \in \CM_N(\IC)$ is an $N\times N$ complex matrix. 
Note that $\psi_N$ induces the isomorphism \eqref{eq:section}
in cohomology. 
Moreover, we have 
\[
\sigma^*(\psi_{N,m-l})^\roof =(-1)^{l+\omega}\psi_{N,l}.
\]
Referring back to \eqref{eq:quasisomBcond}, we see that this
last equation constrains the matrix $M$:
\be\label{eq:sospcases}
\omega = 
\begin{cases} 
0, & \mathrm{if}\ M=M^{tr}\\
1, & \mathrm{if}\ M = -M^{tr}\\
\end{cases}
\ee
The first case corresponds to an $SO(N)$ gauge group, while
the second case corresponds to $Sp(N/2)$ ($N$ even).
This confirms the correlation between the symmetry
of $\psi_N$ and the $SO/Sp$ projection, as we alluded to after
\eqref{eq:psicond}.
 
The infinitesimal deformations of the D-brane are now 
parameterized by a matrix valued field 
\[
\phi={\mathsf C}(\msc^{1,0}+\msc^{0,1}) 
\]
where ${\mathsf C}\in \CM_N(\IC)$ is the $N\times N$ Chan-Paton matrix.
Taking \eqref{eq:genB} into account, invariance under the orientifold
projection yields the following condition on ${\mathsf C}$ 
\be\label{eq:oractionA}
{\mathsf C} = -(-1)^\omega {\mathsf C}^{tr}.
\ee
For $\omega=1$, this condition does not look like the usual one defining
the Lie algebra of $Sp(N/2)$ because we are working in a non-usual basis
of fields, namely $\CC(P(\me_N), \me_N)$. By composing with the
quasi-isomorphism $\psi_N$, we  find the Chan-Paton matrix in
 $\CC(P(\me_N), P(\me_N))$ to be $M\mathsf{C}$. By performing a change of basis
in the space of Chan-Paton indices, we can choose $M$ to be
\[
M = 
\begin{cases} 
\mathrm{I}_N, & \mathrm{if}\ \omega=0\\
i\left(\begin{smallmatrix} & \mathrm{I}_{N/2} \\ - \mathrm{I}_{N/2} & \end{smallmatrix}\right),
& \mathrm{if}\ \omega=1\\
\end{cases}
\]
and so the Chan-Paton matrices satisfy the well-known conditions
\cite{Gimon:1996rq}
\[
\begin{aligned}
(M\mathsf{C})^{tr} & = -(M\mathsf{C}), &\quad\textrm{for } \omega = 0, \\
(M\mathsf{C})^{tr} & = -M(M\mathsf{C})M, &\quad\textrm{for } \omega = 1.
\end{aligned}
\]

The superpotential is determined by the 
$A_\infty$ products \eqref{eq:minmodB} evaluated on 
$\phi$. According to Theorem~\ref{c:20}, the final result is 
obtained by the superpotential of the underlying unprojected 
theory evaluated on invariant field configurations.
Therefore the computations are identical in both cases 
($\omega=0,1$) and the superpotential is essentially determined 
by the $A_\infty$ products of a single D-brane with multiplicity 
$N=1$. 

Proceeding by analogy with \cite{AK}, let us define the cocycles
\[
{\mathsf a}_p \in \check{C}^1(\mfu,\sHom^0(P(\me),\me))\qquad 
{\mathsf b}_{p} \in \check{C}^1(\mfu,\sHom^1(P(\me),\me))
\]
as follows 
\begin{equation}\label{eq:cocyclesA}
\begin{xy} <1em,0em>:
(0,0)*\xybox{\xymatrix@C=8mm@M=1mm{
\co(1) \ar[rr] 
\ar[dd]^{\left(0\right)_{01}} & &
{\begin{matrix} \co^{\oplus 2} \\ \oplus \\ \co(1) \end{matrix}} 
\ar[rr] \ar[dd]^{\left(\begin{smallmatrix} 0 & 0 & 0 \\ 0 & 0 & 0 \\ 0 & 0 
& -x^{-1}y_{2}^{p} \end{smallmatrix}\right)_{01}} & &
{\begin{matrix} \co(1) \\ \oplus \\ \co^{\oplus 2} \end{matrix}} 
\ar[dd]^{\left(0\right)_{01}} \\ \\
{\begin{matrix} \co(-1) \\ \oplus \\ \co^{\oplus 2} \end{matrix}}
\ar[rr] & &
{\begin{matrix} \co^{\oplus 2} \\ \oplus \\ \co(-1) \end{matrix}}
\ar[rr] & & \co(-1)
}},
(-4,-5.5)*{\mathsf{a}_p := }
\end{xy}
\end{equation}
\[
\begin{xy} <1em,0em>:
(0,0)*\xybox{\xymatrix@C=8mm@M=1mm{
\co(1) \ar[rr] \ar[dd]^{\left(\begin{smallmatrix} 0 \\ 0 \\ x^{-1}y_2^p 
\end{smallmatrix}\right)_{01}} & &
{\begin{matrix} \co^{\oplus 2} \\ \oplus \\ \co(1) \end{matrix}} 
\ar[dd]^{\left(\begin{smallmatrix}0 & 0 &  -x^{-1}y_2^p 
\end{smallmatrix}\right)_{01}} \\ \\
{\begin{matrix} \co^{\oplus 2} \\ \oplus \\ \co(-1) \end{matrix}} \ar[rr] & &
\co(-1)
}},
(-4,-5.5)*{\mathsf{b}_p := }
\end{xy} 
\]
One shows by direct computation that they satisfy
the relations 
\be\label{eq:relationsA}
\begin{aligned} 
\msb_p & = Q(\msa_{p-1}) \cr
\msb_p & = {\mathsf c}\star_\psi \msa_p + 
\msa_p \star_\psi {\mathsf c}\cr
\end{aligned}
\ee
Moreover, we have 
\be\label{eq:squareA}
\begin{aligned}
\mscc\star_\psi\mscc & = \msb_{n-2}\cr 
\msb_p \star_\psi \msb_p & =0\cr
\end{aligned} 
\ee 
for any $p$. 
Therefore the computation of the $A_\infty$ products is identical 
to \cite{AK}. We find only one non-trivial product 
\be\label{eq:product} 
{\mathfrak m}_n(\mscc, \ldots, \mscc) = -(-1)^{\frac{n(n-1)}{2}} \msb_0.
\ee
If we further compose with $\mscc$ we obtain
\[
\begin{xy} <1em,0em>:
(0,0)*\xybox{\xymatrix@C=8mm@M=1mm{
& \co(1) \ar[d]^{\left(-x^{-1}\right)_{01}} \\
& \co(-1)\\}},
(-2,-1.7)*{\msb_0 \star_\psi \mathsf{c} := }
\end{xy} 
\]
which is a generator of $\Ext^3(i_* \co_C(-1),i_* \co_C(-1))$.
Therefore we obtain a superpotential of the form
\[ 
W= \frac{(-1)^n}{n+1} \, {\mathsf C}^{n+1} 
\]
where ${\mathsf C}$ satisfies the invariance condition 
\eqref{eq:oractionA}.

\subsection{Local Conifold $O3/O7$ Models} 

In this case, the local Calabi-Yau threefold $X$ is isomorphic 
to the crepant resolution of a conifold singularity,
i.e., the total space of $\co(-1)\oplus \co(-1)\to
\IP^1$.  $X$ can be covered with two coordinate patches 
$(x,y_1,y_2)$, $(w,z_1,z_2)$ with transition functions 
\be\label{eq:extwoA}
\begin{aligned} 
w & = x^{-1}\cr
z_1&=xy_1\cr
z_2&=xy_2.\cr 
\end{aligned}
\ee
The $(-1,-1)$ curve $C$ is given by 
\be\label{eq:extwoB}
x=y_1=y_2=0\qquad w=z_1=z_2=0
\ee 
and the holomorphic involution takes
\be\label{eq:extfourA} 
\begin{aligned} 
(x,y_1,y_2) & \mapsto (-x,-y_1,-y_2) \\
(w,z_1,z_2) & \mapsto (-w,z_1,z_2). 
\end{aligned} 
\ee
In this case we have an $O3$ plane at 
\[ 
x=y_1=y_2=0
\] 
and a noncompact $O7$ plane at $w=0$. 
The invariant D5-brane configurations are of the form 
$\CE_n^{\oplus{N}}$, where 
\be\label{eq:extfourB} 
\CE_n={  {i_*\co_C(-1+n)}}\oplus 
{  {i_*(\sigma^*\co_C(-1-n))}}[1],\quad n\geq 1.
\ee

We have a global Koszul resolution of the structure sheaf $\co_C$ 
\be\label{eq:Koszul} 
\xymatrix@C=8mm@M=1mm{
0 \ar[r] & \co(2) \ar[rr]^{\left(\begin{smallmatrix} -y_{2} \\ y_{1}
    \end{smallmatrix}\right)} && \co(1)^{\oplus 2}
\ar[rr]^{\quad\left(\begin{smallmatrix} y_{1} & y_{2}
    \end{smallmatrix}\right)} && \co \ar[r] & 0
}
\ee
Therefore the locally free resolution of $\CE_n$ is a complex 
$\me_n$ of the form 
\be\label{eq:locfreecpx}
\xymatrix@C=8mm@M=1mm{
\sigma^*\co(1-n) \ar[rr]^{\left(\begin{smallmatrix} 0 \\ y_{1} \\ y_{2}
    \end{smallmatrix}\right)}  & &
{\begin{matrix} \co(1+n) \\ \oplus \\ \sigma^*\co(-n)^{\oplus 2} \end{matrix}} 
\ar[rrr]^{\left(\begin{smallmatrix} -y_{2} & 0 & 0 \\ y_{1} & 0 & 0 \\ 0 &
      y_{2} & -y_{1} 
\end{smallmatrix}\right)} & & &
{\begin{matrix} \co(n)^{\oplus 2} \\ \oplus \\ \sigma^*\co(-1-n) \end{matrix}} 
\ar[rr]^{\left(\begin{smallmatrix} y_{1} & y_{2} & 0 \end{smallmatrix}\right)} 
 & &
\co(-1+n) \\ }
\ee
in which the last term to the right has degree 0, and the last term to the 
left has degree $-3$. 
The quasi-isomorphism 
$\psi \colon\me_n \to P(\me_n) $
is given by 
\be\label{eq:quasi}
\xymatrix@C=8mm@M=1mm{
\sigma^{\ast}\co(1-n) \ar[rr]^{\left(\begin{smallmatrix} 0 \\ y_{1} \\ y_{2} 
\end{smallmatrix}\right)} \ar[dd]^{1} & &
{\begin{matrix} \co(1+n) \\ \oplus \\ \sigma^{\ast}\co(-n)^{\oplus 2} 
\end{matrix}} \ar[rrr]^{\left(\begin{smallmatrix} -y_{2} & 0 & 0 \\ y_{1} & 0
& 0 \\ 0 & y_{2} & -y_{1} \end{smallmatrix}\right)} 
\ar[dd]^{\left(\begin{smallmatrix} & 1 & \\ & & 1 \\ 1 & & \end{smallmatrix}\right)} & & &
{\begin{matrix} \co(n)^{\oplus 2} \\ \oplus \\ \sigma^{\ast}\co(-1-n)
  \end{matrix}} \ar[rr]^{\left(\begin{smallmatrix} y_{1} & y_{2} & 0
    \end{smallmatrix}\right)} \ar[dd]^{\left(\begin{smallmatrix} & & 1 \\ 1 &
      & \\ & 1 & 
\end{smallmatrix}\right)} & &
\co(-1+n) \ar[dd]^{1} \\ \\
\sigma^{\ast}\co(1-n) \ar[rr]^{\left(\begin{smallmatrix} y_{1} \\ y_{2} \\ 0 
\end{smallmatrix}\right)} & &
{\begin{matrix} \sigma^{\ast}\co(-n)^{\oplus 2} \\ \oplus \\ \co(1+n)
  \end{matrix}} 
\ar[rrr]^{\left(\begin{smallmatrix} y_{2} & -y_{1} & 0 \\ 0 & 0 & -y_{2} \\ 0
      & 0 & y_{1} 
\end{smallmatrix}\right)} & & &
{\begin{matrix} \sigma^{\ast}\co(-1-n) \\ \oplus \\ \co(n)^{\oplus 2} 
\end{matrix}} \ar[rr]^{\left(\begin{smallmatrix} 0 & y_{1} & y_{2} 
\end{smallmatrix}\right)} & &
\co(-1+n)
}
\ee
and satisfies $\sigma^*(\psi_{3-l})^\roof = \psi_l$,
that is, the symmetry condition \eqref{eq:psicond} with $\omega = 0$.
The on-shell open string states $\Ext^1_X(\me_n,\me_n)$ are computed by the spectral sequence \eqref{eq:specseq}:
\be\label{eq:spectrum} 
\begin{aligned}
&\Ext^{1}_X(\co_{C}(-1+n), \co_{C}(-1+n))& &=&  &0& \\
&\Ext^{1}_X(\sigma^{\ast}\co_{C}(-1-n)[1], \sigma^{\ast}\co_{C}(-1-n)[1])& &=&  &0& \\
&\Ext^{1}_X(\co_{C}(-1+n), \sigma^{\ast}\co_{C}(-1-n)[1])& &=&
&\mathbb{C}^{4n}& \\
&\Ext^{1}_X(\sigma^{\ast}\co_{C}(-1-n)[1], \co_{C}(-1+n))& &=& 
&\mathbb{C}^{2n+1}, & \\
\end{aligned}
\ee
where in the last two lines we have used the condition $n\geq 1$. 

To compute the superpotential, we work with the cochain model 
$\check{C}(\mfu,{\sHom}(P(\me_n),\me_n))$.
The direct sum of the above Ext groups represents the degree 
1 cohomology of this complex with respect to the differential 
\eqref{eq:cechB}.
The first step is to find explicit representatives for all degree 
1 cohomology classes with well defined transformation properties 
under the orientifold projection. We list all generators below on a case 
by case basis. 

\bigskip
$a)\ \Ext^{1}(\sigma^*\co_{C}(-1-n)[1], \co_{C}(-1+n))$
\bigskip

\noindent 
We have $2n+1$ generators 
$\mathsf{a}_i \in \check{C}^0(\mfu,{\sHom}^1(\CP(\me_n),\me_n)$, $i=0,\ldots 2n$, 
given by 
\begin{equation}\label{eq:genC}
\mathsf{a}_{i} := x^{i} {\mathsf{a}},
\end{equation}
where 
\be\label{eq:genD}
\begin{xy} <1em,0em>:
(0,0)*\xybox{\xymatrix@C=8mm@M=1mm{
\sigma^{\ast}\co(1-n) \ar[rr] 
\ar[dd]^{\left(\begin{smallmatrix} 1 \\ 0 \\ 0 \end{smallmatrix}\right)} & &
{\begin{matrix} \sigma^{\ast}\co(-n)^{\oplus 2} \\ \oplus \\ \co(1+n)
  \end{matrix}} 
\ar[rr] \ar[dd]^{\left(\begin{smallmatrix} 0 & 1 & 0 \\ -1 & 0 & 0 \\ 0 & 0 &
      0 \end{smallmatrix}\right)} & &
{\begin{matrix} \sigma^{\ast}\co(-1-n) \\ \oplus \\ \co(n)^{\oplus 2} 
\end{matrix}} \ar[dd]^{\left(\begin{smallmatrix} 1 & 0 & 0 \end{smallmatrix}\right)} \\ \\
{\begin{matrix} \co(1+n) \\ \oplus \\ \sigma^{\ast}\co(-n)^{\oplus 2} 
\end{matrix}} \ar[rr] & &
{\begin{matrix} \co(n)^{\oplus 2} \\ \oplus \\ \sigma^{\ast}\co(-1-n) 
\end{matrix}} \ar[rr] & &
\co(-1+n)
}},
(-4,-4.5)*{ {\mathsf{a}} := }
\end{xy}
\ee
Note that we have written down the expressions of the generators only 
in the $U_0$ patch.\footnote{ The expressions in the $U_1$ patch can be obtained 
using the transition functions \eqref{eq:extwoA} since the $\mathsf{a}_i$ 
are $\check{\rm C}$ech closed. They will not be needed in the computation.  }
The transformation properties under the orientifold projection are   
\be\label{eq:genE} 
J_\ast P(\mathsf{a}_i) = -(-1)^{i+\omega}\mathsf{a}_i, \qquad 0\leq i \leq 2n.
\ee 

\bigskip
$b)\ \Ext^{1}(\co_{C}(-1+n), \sigma^*\co_{C}(-1-n)[1])$
\bigskip

\noindent 
We have $4n$ generators $\mathsf{b}_i,\mathsf{c}_i\in  
\check{C}^1(\mfu,{\sHom}^0(P(\mf_n),\mf_n)$, $i=1,\ldots,2n$ given by 
\be\label{eq:genF}
\mathsf{b}_{i} := x^{-i} {\mathsf{b}},\qquad 
\mathsf{c}_{i} := x^{-i} {\mathsf{c}}
\ee
where 
\be\label{eq:genG} 
\begin{xy} <1em,0em>:
(0,0)*\xybox{\xymatrix@C=8mm@M=1mm{
{\begin{matrix} \sigma^{\ast}\co(-n)^{\oplus 2} \\ \oplus \\ \co(1+n)
  \end{matrix}} 
\ar[rr] \ar[dd]^{\left(\begin{smallmatrix} 0 & 0 & 0 \\ 0 & 0 & -1 \\ 0 & 0 &
      0 \end{smallmatrix}\right)_{01}} & &
{\begin{matrix} \sigma^{\ast}\co(-1-n) \\ \oplus \\ \co(n)^{\oplus 2} 
\end{matrix}} \ar[dd]^{\left(\begin{smallmatrix} 0 & 0 & 0 \\ 0 & 0 & 0 \\ 0 &
      1 & 0 \end{smallmatrix}\right)_{01}} \\ \\
{\begin{matrix} \co(1+n) \\ \oplus \\ \sigma^{\ast}\co(-n)^{\oplus 2} 
\end{matrix}} \ar[rr] & &
{\begin{matrix} \co(n)^{\oplus 2} \\ \oplus \\ \sigma^{\ast}\co(-1-n) \end{matrix}}
}},
(-4,-4.5)*{ {\mathsf{b}} := }
\end{xy}
\end{equation}
\be\label{eq:genH}
\begin{xy} <1em,0em>:
(0,0)*\xybox{\xymatrix@C=8mm@M=1mm{
{\begin{matrix} \sigma^{\ast}\co(-n)^{\oplus 2} \\ \oplus \\ 
\co(1+n) \end{matrix}} \ar[rr] 
\ar[dd]^{\left(\begin{smallmatrix} 0 & 0 & 0 \\ 0 & 0 & 0 \\ 0 & 0 & -1 
\end{smallmatrix}\right)_{01}} & &
{\begin{matrix} \sigma^{\ast}\co(-1-n) \\ \oplus \\ \co(n)^{\oplus 2} 
\end{matrix}} \ar[dd]^{\left(\begin{smallmatrix} 0 & 0 & 0 \\ 0 & 0 & 0 \\ 0 &
    0 & 1 
\end{smallmatrix}\right)_{01}} \\ \\
{\begin{matrix} \co(1+n) \\ \oplus \\ \sigma^{\ast}\co(-n)^{\oplus 2}
  \end{matrix}} 
\ar[rr] & &
{\begin{matrix} \co(n)^{\oplus 2} \\ \oplus \\ \sigma^{\ast}\co(-1-n) 
\end{matrix}}
}},
(-4,-4.5)*{ {\mathsf{c}} := }
\end{xy}
\end{equation}
The action of the orientifold projection is 
\be\label{eq:oractA}
J_\ast P(\mathsf{b}_i) = (-1)^{i+\omega} \mathsf{b}_i,\qquad
J_\ast P(\mathsf{c}_i) = (-1)^{i+\omega} \mathsf{c}_i.
\ee
For multiplicity $N \geq 1$, we work as in the last subsection, taking the locally free resolution
$\me_{n,N} = \me_n \otimes \IC^N$,
together with the quasi-isomorphism $\psi_N \colon \me_{n,N} \to P(\me_{n,N})$;
$\psi_N = \psi \otimes M$.
Again, $M$ is a symmetric matrix for $\omega = 0$ and  antisymmetric for $\omega = 1$.
A general invariant degree one cocycle $\phi$ will be a linear 
combination 
\be\label{eq:invcocycle} 
\phi = \sum_{i=0}^{2n}\mathsf{A}^i \msa_i+ \sum_{i=1}^{2n}(\mathsf{B}^i \msb_i + \mathsf{C}^i \msc_i)
\ee
where $A^i,B^i,C^i$ are $N\times N$ matrices satisfying 
\be\label{eq:symcond}
(\mathsf{A}^i)^{tr} = -(-1)^{i+\omega} \mathsf{A}^i\qquad 
(\mathsf{B}^i)^{tr} = (-1)^{i+\omega} \mathsf{B}^i \qquad  
(\mathsf{C}^i)^{tr} = (-1)^{i+\omega} \mathsf{C}^i. 
\ee
In the following we will let the indices $i,j,k,\ldots$ run from $0$ to $2n$ 
with the convention $\mathsf{B}^0=\mathsf{C}^0=0$.

The multiplication table of 
the above generators with respect to the product 
\eqref{eq:psiprod} is 
\be\label{eq:multableA} 
\begin{aligned}
{\mathsf{a}}_{i}\star_\psi  {\mathsf{a}}_{j} & ={\mathsf{b}}_{i}\star_\psi{\mathsf{b}}_{j} 
={\mathsf{c}}_{i}\star_\psi {\mathsf{c}}_{j} = 0\cr
{\mathsf{b}}_{i}\star_\psi {\mathsf{c}}_{j} & = {\mathsf{c}}_{i}\star_\psi
{\mathsf{b}}_{j} = 0.\cr
\end{aligned}
\end{equation}
The remaining products are all $Q$-exact:
\[
\begin{aligned}
\msa_i \star_\psi \msb_j & = Q(f_2(\msa_i,\msb_j))\cr
\msb_i\star_\psi \msa_j & = Q(f_2(\msb_i,\msa_j))\cr
\end{aligned}
\] 
as required in \eqref{eq:intcondB}. 
Let us show a sample computation.
\begin{equation}
\begin{xy} <1em,0em>:
(0,0)*\xybox{\xymatrix@C=8mm@M=1mm{
\sigma^{\ast}\co(1-n) \ar[rr] \ar[dd]^{x^{-i+j}\left(\begin{smallmatrix} 0 \\
      -1 \\ 0 
\end{smallmatrix}\right)_{01}} & &
{\begin{matrix} \sigma^{\ast}\co(-n)^{\oplus 2} \\ \oplus \\ \co(1+n)
  \end{matrix}} 
\ar[dd]^{x^{-i+j}\left(\begin{smallmatrix} 0 & 0 & 0 \\ 0 & 0 & 0 \\ 0 & 1 & 0 
\end{smallmatrix}\right)_{01}} \\ \\
{\begin{matrix} \co(1+n) \\ \oplus \\ \sigma^{\ast}\co(-n)^{\oplus 2} 
\end{matrix}} \ar[rr] & &
{\begin{matrix} \co(n)^{\oplus 2} \\ \oplus \\ \sigma^{\ast}\co(-1-n) 
\end{matrix}}
}},
(-7,-2.9)*{\mathsf{b}_{i}\star\mathsf{a}_{j} = }
\end{xy}
\end{equation}

For $j \geq i$,
\begin{equation}
\begin{xy} <1em,0em>:
(0,0)*\xybox{\xymatrix@C=8mm@M=1mm{
\sigma^{\ast}\co(1-n) \ar[rr] \ar[dd]^{x^{-i+j}
\left(\begin{smallmatrix} 0 \\ -1 \\ 0 \end{smallmatrix}\right)_{0}} & &
{\begin{matrix} \sigma^{\ast}\co(-n)^{\oplus 2} \\ \oplus \\ \co(1+n)
  \end{matrix}} 
\ar[dd]^{x^{-i+j}\left(\begin{smallmatrix} 0 & 0 & 0 \\ 0 & 0 & 0 \\ 0 & 1 & 0 
\end{smallmatrix}\right)_{0}} \\ \\
{\begin{matrix} \co(1+n) \\ \oplus \\ \sigma^{\ast}\co(-n)^{\oplus 2} 
\end{matrix}} \ar[rr] & &
{\begin{matrix} \co(n)^{\oplus 2} \\ \oplus \\ \sigma^{\ast}\co(-1-n) 
\end{matrix}}
}},
(-7,-2.9)*{f_{2}(\mathsf{b}_{i}, \mathsf{a}_{j}) = }
\end{xy}
\end{equation}
For $j<i$, 
\begin{equation}
\begin{xy} <1em,0em>:
(0,0)*\xybox{\xymatrix@C=8mm@M=1mm{
\sigma^{\ast}\co(1-n) \ar[rr] \ar[dd]^{(-1)x^{-i+j+1}
\left(\begin{smallmatrix} 0 \\ 1 \\ 0 \end{smallmatrix}\right)_{1}} & &
{\begin{matrix} \sigma^{\ast}\co(-n)^{\oplus 2} \\ \oplus \\ \co(1+n)
  \end{matrix}} 
\ar[dd]^{(-1)x^{-i+j+1}\left(\begin{smallmatrix} 0 & 0 & 0 \\ 0 & 0 & 0 \\ 0 &
      -1 & 0 
\end{smallmatrix}\right)_{1}} \\ \\
{\begin{matrix} \co(1+n) \\ \oplus \\ \sigma^{\ast}\co(-n)^{\oplus 2} 
\end{matrix}} \ar[rr] & &
{\begin{matrix} \co(n)^{\oplus 2} \\ \oplus \\ \sigma^{\ast}\co(-1-n) 
\end{matrix}}
}},
(-7,-2.9)*{f_{2}(\mathsf{b}_{i}, \mathsf{a}_{j}) = }
\end{xy}
\end{equation}
\begin{equation}
\begin{xy} <1em,0em>:
(0,0)*\xybox{\xymatrix@C=8mm@M=1mm{
{\begin{matrix} \sigma^{\ast}\co(-n)^{\oplus 2} \\ \oplus \\ \co(1+n)
  \end{matrix}} 
\ar[rr] \ar[dd]^{x^{i-j}\left(\begin{smallmatrix} 0 & 0 & 0 \\ 0 & 0 & -1 \\ 0
      & 0 & 0 
\end{smallmatrix}\right)_{01}
} & &
{\begin{matrix} \sigma^{\ast}\co(-1-n) \\ \oplus \\ \co(n)^{\oplus 2}
  \end{matrix}} 
\ar[dd]^{x^{i-j}\left(\begin{smallmatrix} 0 & -1 & 0 \end{smallmatrix}\right)_{01}} \\ \\
{\begin{matrix} \co(n)^{\oplus 2} \\ \oplus \\ \sigma^{\ast}\co(-1-n)
  \end{matrix}} \ar[rr]  & &
\co(-1+n)
}},
(-7,-2.9)*{\mathsf{a}_{i}\star\mathsf{b}_{j} = }
\end{xy}
\end{equation}
For $i\geq j$, 
\begin{equation}
\begin{xy} <1em,0em>:
(0,0)*\xybox{\xymatrix@C=8mm@M=1mm{
{\begin{matrix} \sigma^{\ast}\co(-n)^{\oplus 2} \\ \oplus \\ \co(1+n)
  \end{matrix}} 
\ar[rr] \ar[dd]^{x^{i-j}\left(\begin{smallmatrix} 0 & 0 & 0 \\ 0 & 0 & -1 \\ 0
      & 0 & 0 
\end{smallmatrix}\right)_{0}
} & &
{\begin{matrix} \sigma^{\ast}\co(-1-n) \\ \oplus \\ \co(n)^{\oplus 2}
  \end{matrix}} 
\ar[dd]^{x^{i-j}\left(\begin{smallmatrix} 0 & -1 & 0 \end{smallmatrix}\right)_{0}} \\ \\
{\begin{matrix} \co(n)^{\oplus 2} \\ \oplus \\ \sigma^{\ast}\co(-1-n)
  \end{matrix}} \ar[rr]  
& &
\co(-1+n)
}},
(-7,-2.9)*{f_{2}(\mathsf{a}_{i}, \mathsf{b}_{j}) = }
\end{xy}
\end{equation}
For $i<j$, 
\begin{equation}
\begin{xy} <1em,0em>:
(0,0)*\xybox{\xymatrix@C=8mm@M=1mm{
{\begin{matrix} \sigma^{\ast}\co(-n)^{\oplus 2} \\ \oplus \\ \co(1+n) 
\end{matrix}} \ar[rr] \ar[dd]^{x^{i-j+1}
\left(\begin{smallmatrix} 0 & 0 & 0 \\ 0 & 0 & 1 \\ 0 & 0 & 0 \end{smallmatrix}\right)_{1}
} & &
{\begin{matrix} \sigma^{\ast}\co(-1-n) \\ \oplus \\ \co(n)^{\oplus 2} 
\end{matrix}} \ar[dd]^{x^{i-j+1}\left(\begin{smallmatrix} 0 & 1 & 0 
\end{smallmatrix}\right)_{1}} \\ \\
{\begin{matrix} \co(n)^{\oplus 2} \\ \oplus \\ \sigma^{\ast}\co(-1-n) 
\end{matrix}} \ar[rr]  & &
\co(-1+n)
}},
(-7,-2.9)*{f_{2}(\mathsf{a}_{i}, \mathsf{b}_{j}) = }
\end{xy}
\end{equation}
Since all pairwise products of generators are $Q$-exact, it follows
that the obstruction 
$\Pi(B_1(\phi))= \Pi(\phi\star \phi) $
vanishes. Moreover, the second order deformation 
$f_2(\phi)$ is given by 
\be\label{eq:secondorder}
f_2(\phi)=\sum_{i,j} \left(\mathsf{A}^i\mathsf{B}^jf_2(\msa_i,\msb_j) + \mathsf{B}^i\mathsf{A}^jf_2(\msb_i,\msa_j)
+ \mathsf{A}^i\mathsf{C}^jf_2(\msa_i,\msc_j) + \mathsf{C}^i\mathsf{A}^jf_2(\msc_i,\msa_j)\right).
\ee

Following the recursive algorithm discussed in section \ref{s:3} we compute the next obstruction $
\Pi(\phi\star f_2(\phi) + f_2(\phi)\star \phi)$.
For this, we have to compute products of the form 
\[ 
\alpha_i \star f_2(\alpha_j, \alpha_k),\qquad f_2(\alpha_j,\alpha_k) \star 
\alpha_i.
\]
Again we present a sample computation in detail. 
For $i\geq j$,
\begin{equation}
\begin{xy} <1em,0em>:
(0,0)*\xybox{\xymatrix@C=8mm@M=1mm{
\sigma^{\ast}\co(1-n) \ar[rr] \ar[dd]^{x^{i-j+k}\left(\begin{smallmatrix} 0 \\
      -1 \\ 0 
\end{smallmatrix}\right)_{0}} & &
{\begin{matrix} \sigma^{\ast}\co(-n)^{\oplus 2} \\ \oplus \\ \co(1+n)
  \end{matrix}} 
\ar[dd]^{x^{i-j+k}\left(\begin{smallmatrix} 0 & -1 & 0 \end{smallmatrix}\right)_{0}} \\ \\
{\begin{matrix} \co(n)^{\oplus 2} \\ \oplus \\ \sigma^{\ast}\co(-1-n)
  \end{matrix}} 
\ar[rr] & &
\co(-1+n)
}},
(-7,-2.9)*{-\mathsf{a}_{k}\star f_{2}(\mathsf{b}_{j}, \mathsf{a}_{i}) = }
\end{xy}
\end{equation}
and, for $i<j$, 
\begin{equation}
\begin{xy} <1em,0em>:
(0,0)*\xybox{\xymatrix@C=8mm@M=1mm{
\sigma^{\ast}\co(1-n) \ar[rr] \ar[dd]^{(-1)^{n+1}x^{i-j+k-2n+1}
\left(\begin{smallmatrix} 0 \\ 1 \\ 0 \end{smallmatrix}\right)_{1}} & &
{\begin{matrix} \sigma^{\ast}\co(-n)^{\oplus 2} \\ \oplus \\ \co(1+n) 
\end{matrix}} \ar[dd]^{(-1)^{n+2}x^{i-j+k-2n+1}\left(\begin{smallmatrix} 0 & 1
    & 0 
\end{smallmatrix}\right)_{1}} \\ \\
{\begin{matrix} \co(n)^{\oplus 2} \\ \oplus \\ \sigma^{\ast}\co(-1-n) 
\end{matrix}} \ar[rr] & &
\co(-1+n)
}},
(-7,-2.9)*{-\mathsf{a}_{k}\star f_{2}(\mathsf{b}_{j}, \mathsf{a}_{i}) = }
\end{xy}
\end{equation}
For $k\geq j$, 
\begin{equation}
\begin{xy} <1em,0em>:
(0,0)*\xybox{\xymatrix@C=8mm@M=1mm{
\sigma^{\ast}\co(1-n) \ar[rr] \ar[dd]^{x^{i-j+k}
\left(\begin{smallmatrix} 0 \\ 1 \\ 0 \end{smallmatrix}\right)_{0}} & &
{\begin{matrix} \sigma^{\ast}\co(-n)^{\oplus 2} \\ \oplus \\ \co(1+n) 
\end{matrix}} \ar[dd]^{x^{i-j+k}\left(\begin{smallmatrix} 0 & 1 & 0 
\end{smallmatrix}\right)_{0}} \\ \\
{\begin{matrix} \co(n)^{\oplus 2} \\ \oplus \\ 
\sigma^{\ast}\co(-1-n) \end{matrix}} \ar[rr]  & &
\co(-1+n)
}},
(-7,-2.9)*{-f_{2}(\mathsf{a}_{k},\mathsf{b}_{j})\star\mathsf{a}_{i} = }
\end{xy}
\end{equation}
and, for $k<j$, 
\begin{equation}
\begin{xy} <1em,0em>:
(0,0)*\xybox{\xymatrix@C=8mm@M=1mm{
\sigma^{\ast}\co(1-n) \ar[rr] \ar[dd]^{(-1)^{n-1}x^{i-j+k+1-2n}
\left(\begin{smallmatrix} 0 \\ -1 \\ 0 \end{smallmatrix}\right)_{1}} & &
{\begin{matrix} \sigma^{\ast}\co(-n)^{\oplus 2} \\ \oplus \\ \co(1+n)
  \end{matrix}} 
\ar[dd]^{(-1)^{n}x^{i-j+k+1-2n}\left(\begin{smallmatrix} 0 & -1 & 0 
\end{smallmatrix}\right)_{1}} \\ \\
{\begin{matrix} \co(n)^{\oplus 2} \\ \oplus \\ \sigma^{\ast}\co(-1-n) 
\end{matrix}} \ar[rr]  & &
\co(-1+n)
}},
(-7,-2.9)*{-f_{2}(\mathsf{a}_{k}, \mathsf{b}_{j})\star\mathsf{a}_{i} = }
\end{xy}
\end{equation}

Then the third order products are the following. 
For $k<j\leq i$,
\begin{equation}
\begin{xy} <1em,0em>:
(0,0)*\xybox{\xymatrix@C=8mm@M=1mm{
\sigma^{\ast}\co(1-n) \ar[rr] \ar[dd]^{x^{i-j+k}
\left(\begin{smallmatrix} 0 \\ -1 \\ 0 \end{smallmatrix}\right)} & &
{\begin{matrix} \sigma^{\ast}\co(-n)^{\oplus 2} \\ \oplus \\ \co(1+n) 
\end{matrix}} \ar[dd]^{x^{i-j+k}\left(\begin{smallmatrix} 0 & -1 & 0 
\end{smallmatrix}\right)} \\ \\
{\begin{matrix} \co(n)^{\oplus 2} \\ \oplus \\ \sigma^{\ast}\co(-1-n) 
\end{matrix}} \ar[rr] & &
\co(-1+n)
}},
(-7,-2.9)*{\mathfrak{m}_{3}(\mathsf{a}_{k}, \mathsf{b}_{j}, \mathsf{a}_{i}) = }
\end{xy}
\end{equation}
and, for $i < j \leq k$,
\begin{equation}
\begin{xy} <1em,0em>:
(0,0)*\xybox{\xymatrix@C=8mm@M=1mm{
\sigma^{\ast}\co(1-n) \ar[rr] \ar[dd]^{x^{i-j+k}
\left(\begin{smallmatrix} 0 \\ 1 \\ 0 \end{smallmatrix}\right)} & &
{\begin{matrix} \sigma^{\ast}\co(-n)^{\oplus 2} \\ \oplus \\ \co(1+n)
  \end{matrix}} 
\ar[dd]^{x^{i-j+k}\left(\begin{smallmatrix} 0 & 1 & 0 \end{smallmatrix}\right)} \\ \\
{\begin{matrix} \co(n)^{\oplus 2} \\ \oplus \\ \sigma^{\ast}\co(-1-n)
  \end{matrix}} 
\ar[rr] & &
\co(-1+n)
}},
(-7,-2.9)*{\mathfrak{m}_{3}(\mathsf{a}_{k}, \mathsf{b}_{j}, \mathsf{a}_{i}) = }
\end{xy}
\end{equation}

According to \cite{AK}, the corresponding terms in the superpotential 
can be obtained by taking products of the form 
$\mathfrak{m}_2(\mathfrak{m}_3(\alpha_i,\alpha_j,\alpha_k),\alpha_l) $,
which take values in $\Ext^3(\CE_n, \CE_n)$. 
For $k < j \leq i$ and $i-j+k-l=-1$ we have 
\begin{equation}\label{eq:termA}
\begin{xy} <1em,0em>:
(0,0)*\xybox{\xymatrix@C=8mm@M=1mm{
{\begin{matrix} \sigma^{\ast}\co(-n)^{\oplus 2} \\ \oplus \\ \co(1+n) 
\end{matrix}} \ar[dd]^{x^{i-j+k-l}\left(\begin{smallmatrix} 0 & 0 & 1 
\end{smallmatrix}\right)_{01}} \\ \\
\co(-1+n)
}},
(-7,-2.9)*{\mathfrak{m}_{2}(\mathfrak{m}_{3}(\mathsf{a}_{k}, \mathsf{b}_{j}, 
\mathsf{a}_{i}),  \mathsf{c}_{l})= }
\end{xy}
\end{equation}
The expression obtained in the right hand side of  
equation \eqref{eq:termA} is a generator for 
\be\label{eq:exthree}
\Ext^3(\sigma^\ast\co_C(-1-n)[1],\sigma^\ast\co_C(-1-n)[1]) = \IC.
\ee
For $i < j \leq k$ and $i-j+k-l=-1$,
\begin{equation}\label{eq:termB}
\begin{xy} <1em,0em>:
(0,0)*\xybox{\xymatrix@C=8mm@M=1mm{
{\begin{matrix} \sigma^{\ast}\co(-n)^{\oplus 2} \\ \oplus \\ \co(1+n) 
\end{matrix}} \ar[dd]^{x^{i-j+k-l}\left(\begin{smallmatrix} 0 & 0 & -1 
\end{smallmatrix}\right)_{01}} \\ \\
\co(-1+n)
}},
(-7,-2.9)*{\mathfrak{m}_{2}(\mathfrak{m}_{3}(\mathsf{a}_{k}, \mathsf{b}_{j}, 
\mathsf{a}_{i}),  \mathsf{c}_{l})= }
\end{xy}
\end{equation}
Note that the expression in the right hand side of  
\eqref{eq:termB} is the same generator of 
\eqref{eq:exthree} multiplied by $(-1)$. 
The first product \eqref{eq:termA} gives rise to 
superpotential terms of the form 
\[ 
\mathrm{Tr}(C^lA^kB^jA^i)
\]
with 
\[
(i+k)-(j+l)=-1,\qquad k<j\leq i.
\]
The second product \eqref{eq:termB} gives rise to 
terms in the superpotential of the form 
\[
-\mathrm{Tr}(C^lA^kB^jA^i)
\]
with 
\[
(i+k)-(j+l)=-1,\qquad i<j\leq k.
\]
If we consider the case $n=1$ for simplicity, the superpotential 
interactions resulting from these two products are 
\be\label{eq:supintB} 
\begin{aligned}
W=\mathrm{Tr}(& C^1A^0B^1A^1-C^1A^1B^1A^0+C^2A^0B^1A^2-C^2A^2B^1A^0\cr
& +C^1A^0B^2A^2-C^1A^2B^2A^0+C^2A^1B^2A^2-C^2A^2B^2A^1).\cr
\end{aligned}
\ee

\begin{appendix}
\section{An alternative derivation}\label{s:ss} 

In this appendix we give an alternative derivation of Lemma~\ref{c:1}. This approach relies on one of the most powerful results in algebraic geometry, namely Grothendieck duality. Let us start out by recalling the latter. Consider $f\colon X\to Y$ to be a proper morphism of smooth varieties\footnote{
The Grothendieck duality applies to more general schemes than varieties, but we limit ourselves to the cases considered in this paper.}. Choose $\cF\in D^b(X)$ and $\cG\in D^b(Y )$ to be objects in the corresponding bounded derived categories. Then one has the following isomorphism (see, e.g., III.11.1 of \cite{Hart:dC}):
\begin{equation}\label{a:1}
\mathbf{R}f_* \mathbf{R}\!\sHom_X (\cF,\, f^! \cG)\cong
\mathbf{R}\!\sHom_Y(\mathbf{R}f_* \cF,\,  \cG).
\end{equation}

Now it is true that $f^!$ in general is a complicated functor, in particular it is \emph{not} the total derived functor of a classical functor, i.e., a functor between the category of coherent sheaves, but in our context it will have a very simple form.

The original problem that lead to Lemma~\ref{c:1} was to determine the derived dual, a.k.a, Verdier dual, of a torsion sheaf. Let  $i\colon E\to X$ be the embedding of a codimension $d$ subvariety $E$ into a smooth variety $X$, and let $V$ be a vector bundle on  $E$. We want to determine $ \mathbf{R}\!\sHom_X (i_*V,\co_X)$. Using (\ref{a:1}) we have
\begin{equation}
\mathbf{R}\!\sHom_X (i_*V,\co_X)\cong i_*\mathbf{R}\!\sHom_E(V,i^! \co_X),
\end{equation}
where we used the fact that the higher direct images of $i$ vanish. Furthermore, since $V$ is locally free, we have that 
\begin{equation}
\mathbf{R}\!\sHom_E(V,i^! \co_X)= \mathbf{R}\!\sHom_E(\co_E,V^\roof\otimes i^! \co_X)=V^\roof\otimes i^! \co_X,
\end{equation}
where $V^\roof$ is the dual of $V$ on  $E$, rather than on $X$. On the other hand, for an embedding
\begin{equation}
i^! \co_X=K_{E/X}[-d]\,,
\end{equation}
where $K_{E/X}$ is the relative canonical bundle. 
Now if we assume that the ambient space $X$ is a Calabi-Yau variety, then $K_{E/X}=K_E$. We can summarize this 
\begin{prop}\label{pa1}
For the embedding $i\colon E\to X$ of a codimension $d$ subvariety $E$ in a smooth Calabi-Yau variety $X$, and a vector bundle $V$ on  $E$ we have that 
\begin{equation}
\mathbf{R}\!\sHom_X (i_*V,\co_X)\cong i_*\left(V^\roof\otimes  K_E\right)[-d].
\end{equation}
\end{prop}

\end{appendix}


\providecommand{\href}[2]{#2}\begingroup\raggedright\endgroup

\end{document}